\documentclass[prb,twocolumn,showpacs,superscriptaddress,amsmath,amssymb,showkeys]{revtex4-1}
\usepackage{natbib}
\usepackage{graphicx}         
\usepackage{dcolumn}          
\usepackage{bm}               
\usepackage{lipsum}
\usepackage{upgreek}
\usepackage{booktabs} 				
\usepackage{tabularx}
\usepackage{array}
\usepackage[colorlinks=true,urlcolor=blue,breaklinks=true]{hyperref}
\usepackage{xcolor}
\usepackage{colortbl}
\usepackage{multirow}
\usepackage[version=3]{mhchem}
\usepackage{mathtools}

\begin{document}

\title{Te-doped selective-area grown InAs nanowires for superconducting hybrid devices}

\author{Pujitha Perla}
\thanks{These two authors contributed equally }
\affiliation{Peter Gr\"unberg Institut (PGI-9), Forschungszentrum J\"ulich, 52425 J\"ulich, Germany}
\affiliation{JARA-Fundamentals of Future Information Technology, J\"ulich-Aachen Research Alliance, Forschungszentrum J\"ulich and RWTH Aachen University, Germany}

\author{Anton Faustmann} 
\thanks{These two authors contributed equally }
\affiliation{Peter Gr\"unberg Institut (PGI-9), Forschungszentrum J\"ulich, 52425 J\"ulich, Germany} \affiliation{JARA-Fundamentals of Future Information Technology, J\"ulich-Aachen Research Alliance, Forschungszentrum J\"ulich and RWTH Aachen University, Germany}

\author{Sebastian K\"olling}
\affiliation{Department  of  Engineering  Physics, \'{E}cole  Polytechnique  de Montr\'{e}al, C.P.  6079,  Succ.   Centre-Ville,  Montr\'{e}al,  Qu\'{e}bec,  Canada  H3C  3A7}

\author{Patrick Zellekens}
\affiliation{Peter Gr\"unberg Institut (PGI-9), Forschungszentrum J\"ulich, 52425 J\"ulich, Germany}
\affiliation{JARA-Fundamentals of Future Information Technology, J\"ulich-Aachen Research Alliance, Forschungszentrum J\"ulich and RWTH Aachen University, Germany}

\author{Russell Deacon} \affiliation{RIKEN Center for Emergent Matter Science and Advanced Device Laboratory, 351-0198 Saitama, Japan}

\author{H. Aruni Fonseka}
\affiliation{Department of Physics, University of Warwick, Coventry
CV4 7AL, UK}

\author{Jonas K\"olzer} \affiliation{Peter Gr\"unberg Institut (PGI-9), Forschungszentrum J\"ulich, 52425 J\"ulich, Germany} \affiliation{JARA-Fundamentals of Future Information Technology, J\"ulich-Aachen Research Alliance, Forschungszentrum J\"ulich and RWTH Aachen University, Germany}

\author{Yuki Sato} \affiliation{Peter Gr\"unberg Institut (PGI-9), Forschungszentrum J\"ulich, 52425 J\"ulich, Germany} \affiliation{JARA-Fundamentals of Future Information Technology, J\"ulich-Aachen Research Alliance, Forschungszentrum J\"ulich and RWTH Aachen University, Germany}
\affiliation{Faculty of Science and Engineering Department of Electronics, Doshisha University, Doshisha University, Kyotanabe, Kyoto 610-0321, Japan}

\author{Ana M. Sanchez} \affiliation{Department of Physics, University of Warwick, Coventry CV4 7AL, UK}

\author{Oussama Moutanabbir} \affiliation{Department  of  Engineering  Physics, \'{E}cole  Polytechnique  de Montr\'{e}al, C.P.  6079,  Succ.   Centre-Ville,  Montr\'{e}al,  Qu\'{e}bec,  Canada  H3C  3A7}

\author{Koji Ishibashi} \affiliation{RIKEN Center for Emergent Matter Science and Advanced Device Laboratory, 351-0198 Saitama, Japan}

\author{Detlev Gr\"utzmacher} \affiliation{Peter Gr\"unberg Institut (PGI-9), Forschungszentrum J\"ulich, 52425 J\"ulich, Germany} \affiliation{Peter Gr\"unberg Institut (PGI-10), Forschungszentrum J\"ulich, 52425 J\"ulich, Germany} \affiliation{JARA-Fundamentals of Future Information Technology, J\"ulich-Aachen Research Alliance, Forschungszentrum J\"ulich and RWTH Aachen University, Germany}

\author{Mihail Ion Lepsa} \affiliation{Peter Gr\"unberg Institut (PGI-10), Forschungszentrum J\"ulich, 52425 J\"ulich, Germany} \affiliation{JARA-Fundamentals of Future Information Technology, J\"ulich-Aachen Research Alliance, Forschungszentrum J\"ulich and RWTH Aachen University, Germany}

\author{Thomas~Sch\"apers} \email{th.schaepers@fz-juelich.de} \affiliation{Peter Gr\"unberg Institut (PGI-9), Forschungszentrum J\"ulich, 52425 J\"ulich, Germany} \affiliation{JARA-Fundamentals of Future Information Technology, J\"ulich-Aachen Research Alliance, Forschungszentrum J\"ulich and RWTH Aachen University, Germany}

\date{\today}

\begin{abstract}
Semiconductor nanowires have emerged as versatile components in superconducting hybrid devices for Majorana physics and quantum computing. The transport properties of nanowires can be tuned either by field-effect or doping. We investigated a series of InAs nanowires which conductivity has been modified by $n$-type doping using tellurium. In addition to electron microscopy studies, the wires were also examined with atomic probe tomography to obtain information about the local incorporation of Te atoms. It was found that the Te atoms mainly accumulate in the core of the nanowire and at the corners of the \{110\} side facets. The efficiency of $n$-type doping was also confirmed by transport measurements. As a demonstrator hybrid device, a Josephson junction was fabricated using a nanowire as a weak link. The corresponding measurements showed a clear increase of the critical current with increase of the dopant concentration.
\end{abstract}
\maketitle

\section{Introduction}

InAs nanowires have proven to be suitable consituents for various electronic devices\cite{Dayeh07,Tomioka07a,Thelander08,Burke15} and  also for studying fundamental quantum physics phenomena in nanoscale structures.\cite{Fasth05a,Fuhrer07a,Heedt16b,Heedt17,Iorio18,Bordoloi20} Recently, InAs nanowires attracted  attention by their integration in semiconductor-superconductor hybrid structures. A prominent example is the effort regarding the realization of Majorana fermions for topological quantum computing.\cite{Mourik12,Aguado17} In addition, InAs nanowires can serve as a weak link in gate-controlled Josephson junctions\cite{Doh05,Guenel12} for superconducting qubits such as transmons\cite{Larsen15,deLange15} or Andreev level qubits.\cite{Zazunov03} 

Due to the Fermi level pinning within the conduction band in InAs, a surface accumulation layer  forms naturally,\cite{Wieder74} ensuring that the nanowire is conductive even without doping. In some cases, however, the ability to control conductivity through doping is desired, e.g. for adjusting the channel conductance in field-effect transistors. Nanowire conductivity has also a great impact on the critical current of Josephson junctions based on InAs nanowires\cite{Guenel12} and therefore it is a crucial design parameter for the aforementioned qubit circuits.

In most cases, Si is employed as an $n$-type dopant in InAs nanowires.\cite{Sladek10a,Wirths11,Ghoneim12,Dimakis13,Park15} 
Using Si doping, densities in the order of $10^{19}$\,cm$^{-3}$ are achieved.\cite{Wirths11,Ghoneim12} However,  apart from the increase of the carrier concentration in the nanowire, Si doping affects the growth kinetics as well as the nanowire dimensions.\cite{Wirths11,Dimakis13} In GaAs, Te is known as a very effective $n$-type dopant.\cite{Sankaran80,Kamp94,Bennett03} As a group VI element Te shows no amphoteric behavior as it is the case for Si in GaAs. As a matter of fact, Te doping already found its applications in $n$-type doped GaAs nanowires.\cite{Wallentin11a,Orru16,Goktas18,Hakkarainen19} Recently, it was demonstrated that Te is also a very efficient dopant in InAs nanowires.\cite{Guesken19}

To explore the suitability of Te-doped nanowires in superconductor/nanowire hybrid structures, we have grown a series of nanowires with different doping using selective-area molecular beam epitaxy (MBE). The structural properties are investigated by electron microscopy. Detailed information about the dopant distribution in the nanowires is obtained by atom probe tomography (APT). Electrical transport measurements are performed both at room temperature and at 4\,K to determine the doping efficiency. Finally, the properties of the doped nanowires for superconducting hybrid structures are investigated by fabricating and measuring nanowire-based Josephson junctions.

\section {Growth and Device Fabrication} 

The InAs nanowires were grown by MBE using selective area growth on a Si(111) wafer covered by a 20-nm-thick thermally oxidized SiO$_2$ layer.\cite{Koblmueller10,Perla21} By means of electron beam lithography, hole arrays with a pitch of 1\,$\mu$m and hole diameter of 80\,nm are patterned. The holes are etched by reactive ion etching to a depth of roughly 16\,nm followed by cleaning in piranha solution (H$_2$SO$_4$:H$_2$O$_2 = 3:1$). Immediately before growth, an etching step of 60\,s in HF is performed in order to remove the remaining SiO$_2$ in the holes without damaging the Si(111) surface below.

The InAs nanowires are grown without any catalyst via the vapor-solid mechanism \,$^\circ$C.\cite{Rieger13a} The tellurium is supplied by a GaTe cell. For the first $10\,$min, a substrate temperature of 480\,$^\circ$C, an In growth rate of $0.08\,\mu$m/h and an As beam equivalent pressure (BEP) of $4 \times 10^{-5}$\,mbar are used  to sustain the nanowire self-seeding. The nanowire growth is then proceeded with a lower substrate temperature of $460\,^\circ$C and In growth rate of $0.03\,\mu$m/h. For growth runs A to D the As$_4$ BEP varied between $3-3.5 \times 10^{-5}$\,mbar for $3.5$\,h depending on the Te doping concentration to avoid changes in the morphology of the nanowires.\cite{Guesken19} 
Here, the Te concentration has been varied between $5 \times 10^{17}$ and $1 \times 10^{19}$\,cm$^{-3}$ by varying the GaTe cell temperature between $420\,^\circ$C and $497\,^\circ$C based on calibrations conducted on Te-doped GaAs layers via Hall-measurements. In addition, two samples were grown with a lower As$_4$ BEP of $2.5 \times 10^{-5}$\,torr (growth runs E and F) with a nominal doping of $5 \times 10^{18}$ and $2.5 \times 10^{19}$\,cm$^{-3}$, respectively. The parameters of the Te-doped InAs nanowires are summarized in table~\ref{tab:InAs-samples}.
\begin{table*}
\caption{Growth runs for InAs nanowires: growth run, nominal doping concentration, GeTe cell temperature, As$_4$ BEP, length, diameter, resistivity, and carrier concentration from transport measurements at room temperature. \label{tab:InAs-samples}}
\begin{ruledtabular}
\begin{tabular}{llllllll}
    growth run & Te-doping & GeTe cell &  As$_4$ BEP & length & diameter & $\rho$ & $n_\mathrm{3D}$ \\
    & ($\mathrm{cm}^{-3}$) & ($^\circ$C) & (torr)   & ($\mu$m)& (nm) & ($\Omega\mathrm{cm}$) &($\mathrm{cm}^{-3}$)\\
    \hline
   A 
   & $5 \times 10^{17}$ & 420 & $3 \times 10^{-5}$& $4 - 5$ & $100 \pm 5$ & $18.4 \times 10^{-3}$& $(0.75 \pm 0.23) \times 10^{18}$\\
   B 
   & $1 \times 10^{18}$ & 437 & $3 \times 10^{-5}$& $5 - 6.5$ & $103 \pm 5$ & n.a. & $(1.15 \pm 0.80) \times 10^{18}$\\
   C 
   & $5 \times 10^{18}$ & 478 & $3.5 \times 10^{-5}$& $3.5 - 4.5$  & $111 \pm 5$ & $4.3 \times 10^{-3}$ & $(2.76 \pm 2.50) \times 10^{18}$\\
   D 
   & $1 \times 10^{19}$ & 497 & $3.5 \times 10^{-5}$ & $4 - 5$ & $113 \pm 3$ & $2.7 \times 10^{-3}$ & $(9.04 \pm 2.29) \times 10^{18}$\\
   E 
   & $5 \times 10^{18}$ & 478 & $2.5 \times 10^{-5}$& $3.5 - 5$ & $121 \pm 7$ & $10.0\times 10^{-3}$ & $(1.27 \pm 0.45) \times 10^{18}$\\
   F 
   & $2.5 \times 10^{19} $ & 520 &$2.5 \times 10^{-5}$& $3.5 - 5$ & $140 \pm 8$ & $1.2 \times 10^{-3}$& $(5.86 \pm 2.21) \times 10^{18}$\\
  \end{tabular}
\end{ruledtabular}
\label{tab:nw_params}
\end{table*}

\begin{figure}[!ht]
	\centering
\includegraphics[width=0.90\columnwidth]{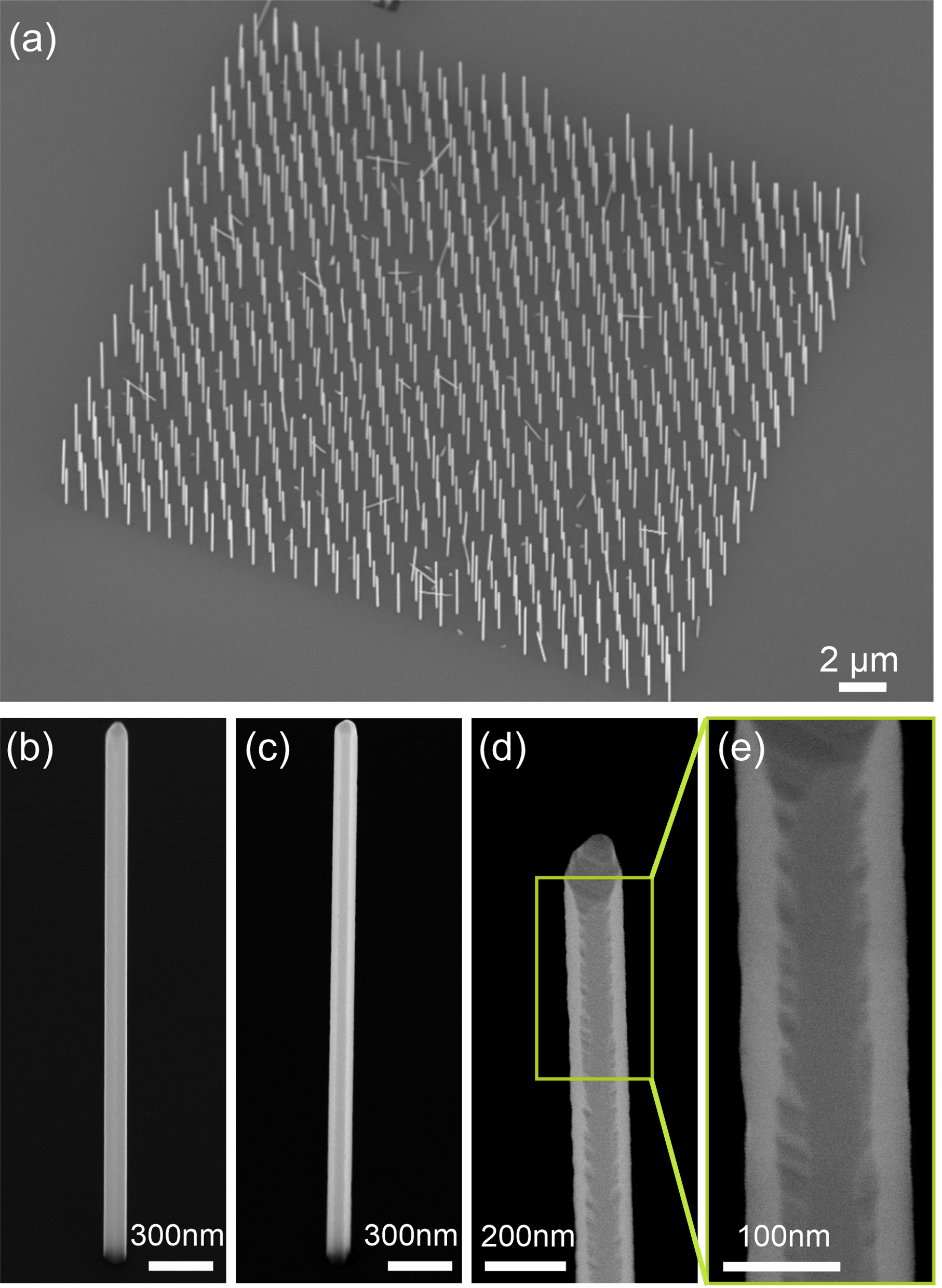}
	\caption[]{(a) Scanning electron micrograph of selectively grown Te-doped InAs nanowires with a nominal doping concentration of $1 \times 10 ^{18}$\,cm$^{-3}$. The nanowires grow in 80-nm-wide holes with 1 $\mu$m pitch. Close-up of nanowires with different Te-concentrations: (b) $1 \times 10 ^{18}$\,cm$^{-3}$ (growth run B), (c) $1 \times 10 ^{19}$\,cm$^{-3}$ (growth run D), (d) $2.5 \times 10 ^{19}$\,cm$^{-3}$ (growth run F), respectively. (e) Zoom-in of (d) showing additional facets forming an irregular dodecagonal shape.}
	\label{fig:SEM-growth}
\end{figure}

 For the APT measurements discussed below, we also prepared InAs nanowires with different doping concentrations, which are covered by a 50-nm-thick GaSb shell (cf. Table~\ref{tab:InAs-GaSb-samples}). The GaSb shell ensures that all atoms of the cross section of the InAs nanowire are gathered by the APT detector.
\begin{table}
\caption{Growth runs for InAs/GaSb core/shell nanowires with their corresponding nominal doping concentration and the doping concentration determined by APT.  
\label{tab:InAs-GaSb-samples} 
} 
\begin{ruledtabular}
\begin{tabular}{lll}
    Growth run & Te-doping & Te (APT) \\
    & ($\mathrm{cm}^{-3}$) & ($\mathrm{cm}^{-3}$) \\
    \hline
      H 
      &  $1  \times 10^{18} $ &$(1.47 - 1.92) \times 10^{18}$ \\
  I 
  &  $7.5  \times 10^{18} $ &$(2.45 - 4.95) \times 10^{18}$ \\
   J 
   &  $2.5 \times 10^{19} $ & $(0.39 - 1.38) \times 10^{19}$ \\
  \end{tabular}
\end{ruledtabular}
\end{table}

Specific devices with ohmic contacts have been fabricated for electrical characterisation of InAs nanowires with varying Te-doping levels from growth runs A-F. The nanowires have been transferred mechanically using a clean paper tip to highly doped Si substrates covered by a  200\,nm thick thermal SiO$_2$ layer in order to provide a global back gate. On top of the SiO$_2$ layer, metallic contact pads and alignment markers have been placed using optical lithography and lift-off processes. Electrical contacts were defined in a four-terminal configuration by electron beam lithography. The contacts consisted of non-alloyed Ti/Au metal layers, 80\,nm and 50\,nm thickness respectively, deposited by electron beam evaporation. An Ar$^+$ sputtering step of 90\,s was included before metallization in order to remove the native oxide and to provide a clean semiconductor surface.

The shunted Josephson junctions were processed with nanowires from growth runs A, B, and D. The AuGe shunt resistor is defined by electron beam lithography. It consists of a 10-nm-thick, 1-$\mu$m-wide and 7-$\mu$m-long AuGe stripe with a resistance  $R_\mathrm{shunt}=80-140\,\Omega$. The InAs nanowires are transferred individually onto a Si substrate containing a 5$\,$nm/10$\,$nm thick Ti/Pt gate pad covered with a 3$\,$nm/12$\,$nm thick stack of Al$_2$O$_3$/HfO$_2$. The  NbTi electrodes that connect the nanowire with the shunt resistor and the surrounding TiN circuit are fabricated by means of Ar ion milling ($\sim 180$\,s) and the subsequent sputter deposition of 80$\,$nm NbTi via DC magnetron. Here, the average junction length is in the range of 100$\,$nm and mainly limited by the e-beam lithography. Junctions were fabricated in the center of the transferred nanowires.

\section{Results and Discussion}

\subsection{Nanowire structure} 

Figure~\ref{fig:SEM-growth} (a) shows a scanning electron microscopy (SEM) image of an array of Te-doped ($1 \times 10^{18}$\,cm$^{-3}$) nanowires (growth run B). The yield is about 95\%. The image confirms the uniformity of length and diameter. In Figs.~\ref{fig:SEM-growth} (b) and (c), close-ups of a nanowire from growth runs B and D are shown. Up to a Te concentration of $1 \times 10^{-19}$\,cm$^{-3}$ the nanowires show a hexagonal cross section with $\{$110$\}$ facets as well as a comparable diameter of around 110\,nm. However, for the nanowires with the highest Te concentration of $2.5 \times 10^{-19}$\,cm$^{-3}$ we observe that diameter increased visibly. Furthermore, as can be clearly seen in Figs.~\ref{fig:SEM-growth} (d) and (e), the sidewalls of the nanowire developed additional \{112\} facets, forming an irregular dodecagonal shape.

Tellurium atoms have a surfactant effect accumulating on the side facets and decreasing the diffusion length of the host indium atoms.\cite{Wixom04,Safar97,Neves98}  As a result, an increase of the nanowire diameter  is expected, which was observed also in our experiments (cf. table~\ref{tab:InAs-samples}). A similar behavior has Sb in InAs(Sb) nanowires, where with increasing Sb supply the nanowire radius increases accordingly.\cite{Anyebe15,Potts16}

\subsection{Transmissions Electron Microscopy}

\begin{figure}[!ht]
	\centering
\includegraphics[width=0.95\columnwidth]{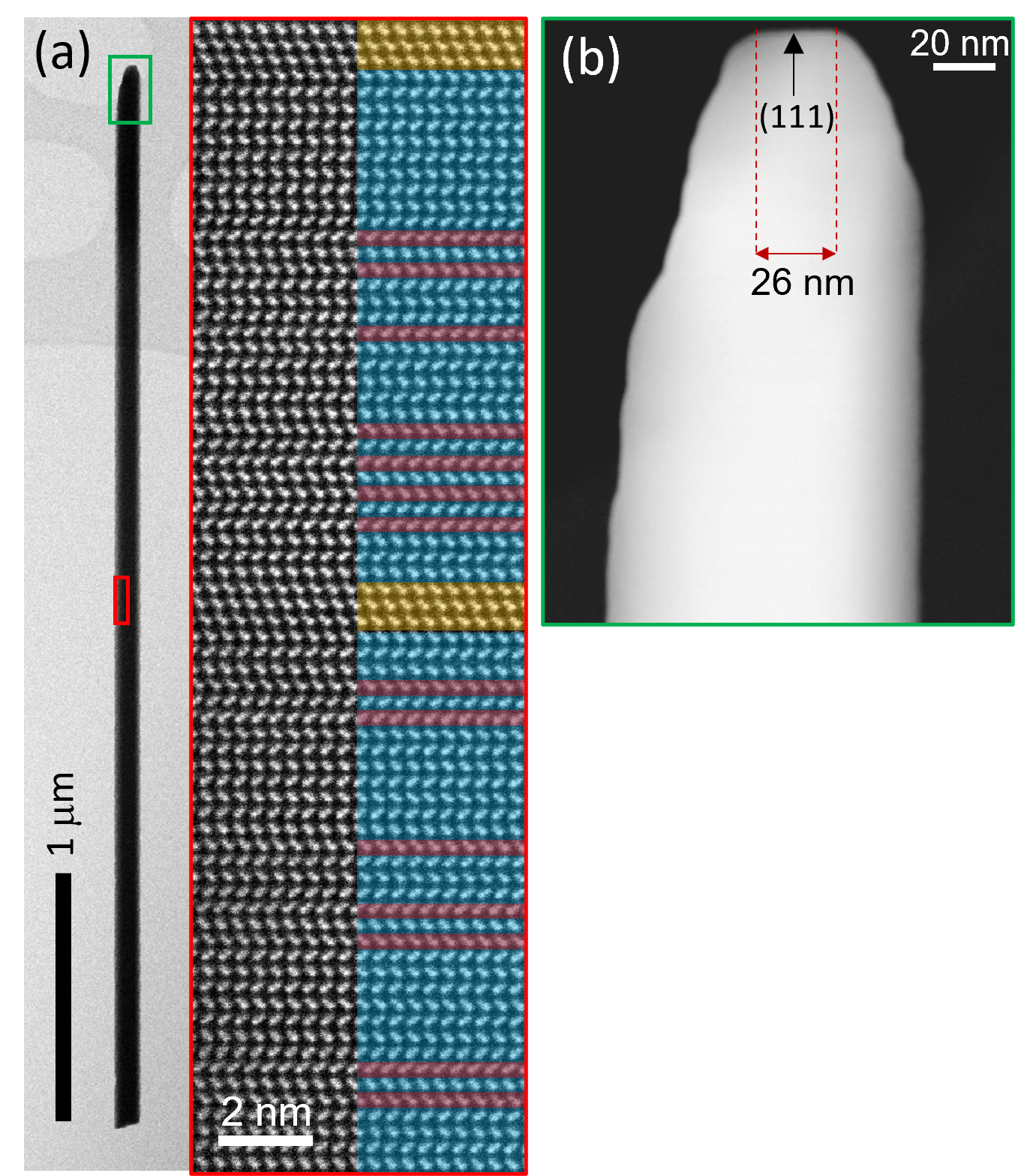}
\caption[]{(a) A high magnification annular dark field (ADF) image of a nanowire from growth run C showing the typical crystal structure of the Te doped nanowires. WZ regions containing stacking faults are marked in blue and red, respectively, and ZB inclusions are marked in yellow. The region used for the magnified image is indicated in the inset bright field (BF) low magnification image. (b) Higher magnification ADF image of the tip region. It appears to consist of multiple facets with the top (111) facet being smaller in diameter ($\approx 26$\,nm) compared to the nanowire.}
	\label{fig:TEM}
\end{figure}

The crystal structure of the nanowires from all the growths is mainly formed by wurtzite (WZ) region containing inclusions of zinc blende (ZB) segments and twining planes, as shown in Fig.~\ref{fig:TEM} (a) for a nanowire from growth run C. In contrast to the previous study on randomly positioned Te-doped InAs nanowires,\cite{Guesken19}  in these nanowires, no clear clear change in crystal structure was observed with the increase of the Te doping level.
Figure~\ref{fig:TEM} (b) shows a higher magnification image of the tip of the nanowire. Assuming similar facets are maintained during growth, the top (111) growth facet is of smaller diameter than the rest of the nanowire. Therefore, it can be safely assumed that the nanowire growth takes place by adatom deposition on multiple facets on the tip region.

\subsection{Atomic Probe Tomography}

\begin{figure*}[!ht]
	\centering
\includegraphics[width=0.85\textwidth]{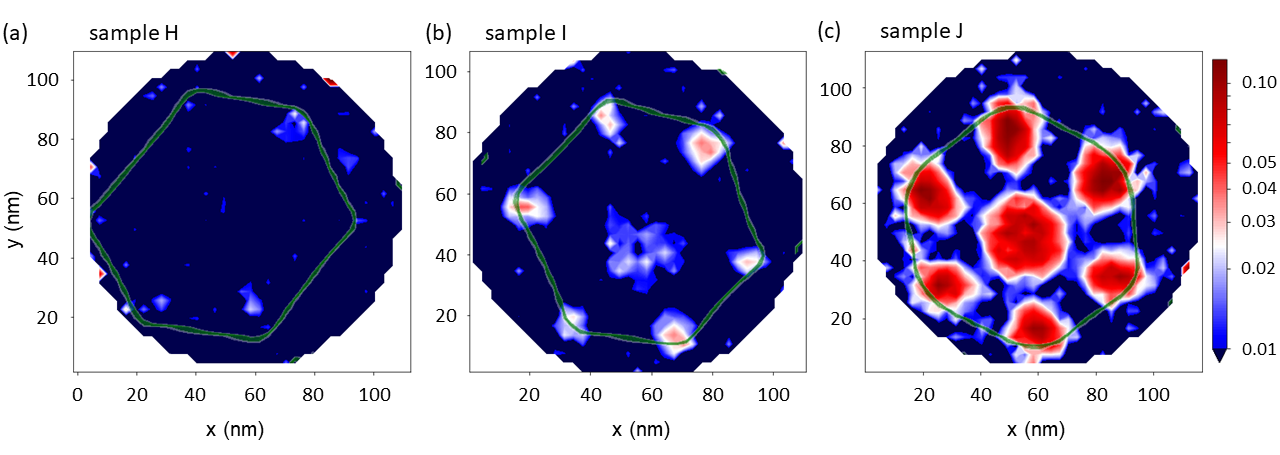}
	\caption[]{Two-dimensional map of the Te concentration determined by APT from samples of growth runs H (a), I (b), and J(c).}
	\label{fig:APT-cross}
\end{figure*}

\begin{figure}[!ht]
	\centering
\includegraphics[width=0.98\columnwidth]{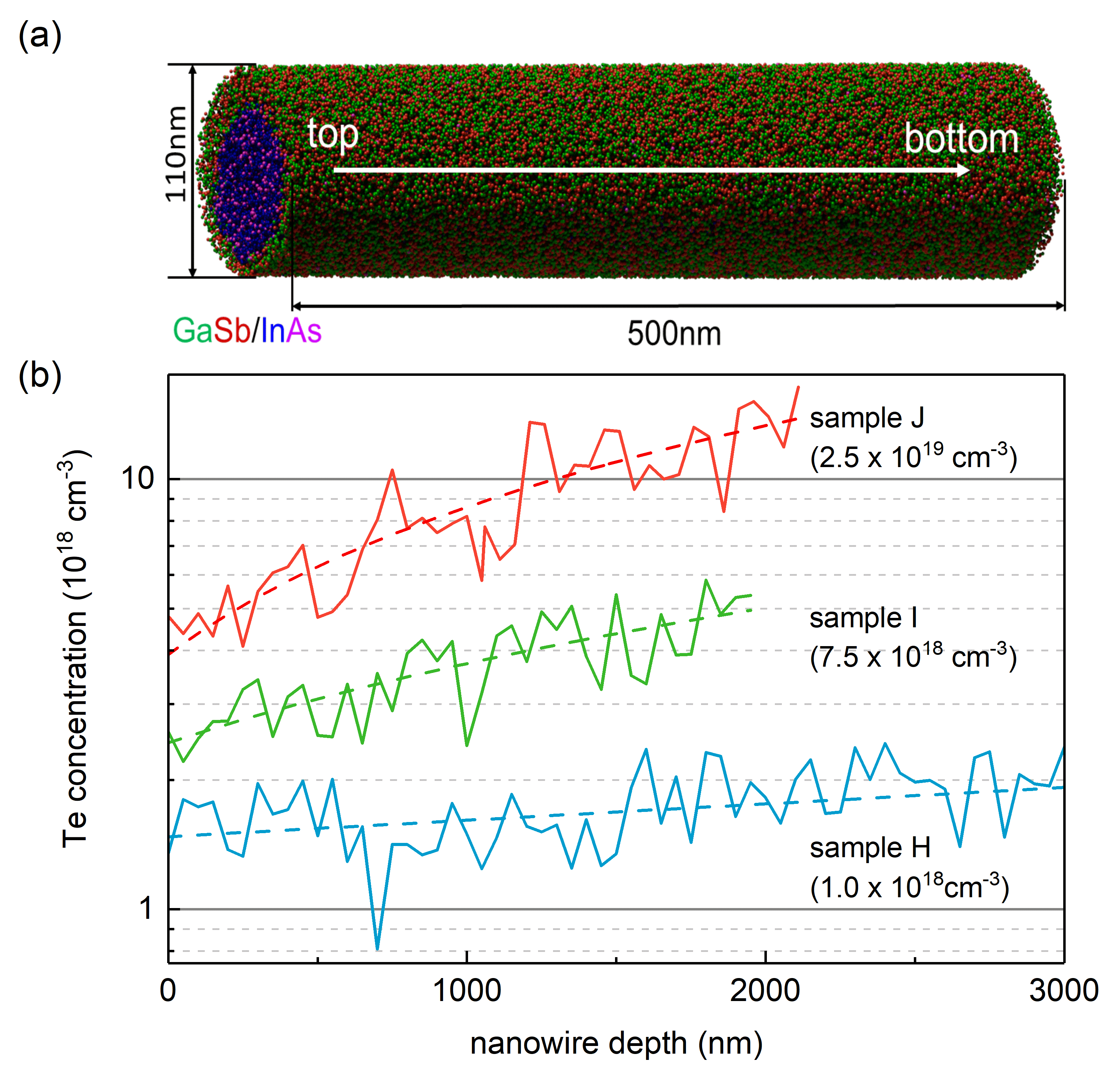}
	\caption[]{(a) APT 3D map of an exemplary InAs/GaSb core/shell nanowire. The elements are displayed with different colors.  (b) Axial atomic Te concentration profiles of samples from growth runs H, I, and J. The nominal Te-doping concentration is given in brackets. }
	\label{fig:APT-conc}
\end{figure}

\begin{figure*}[!ht]
	\centering
\includegraphics[width=0.75\textwidth]{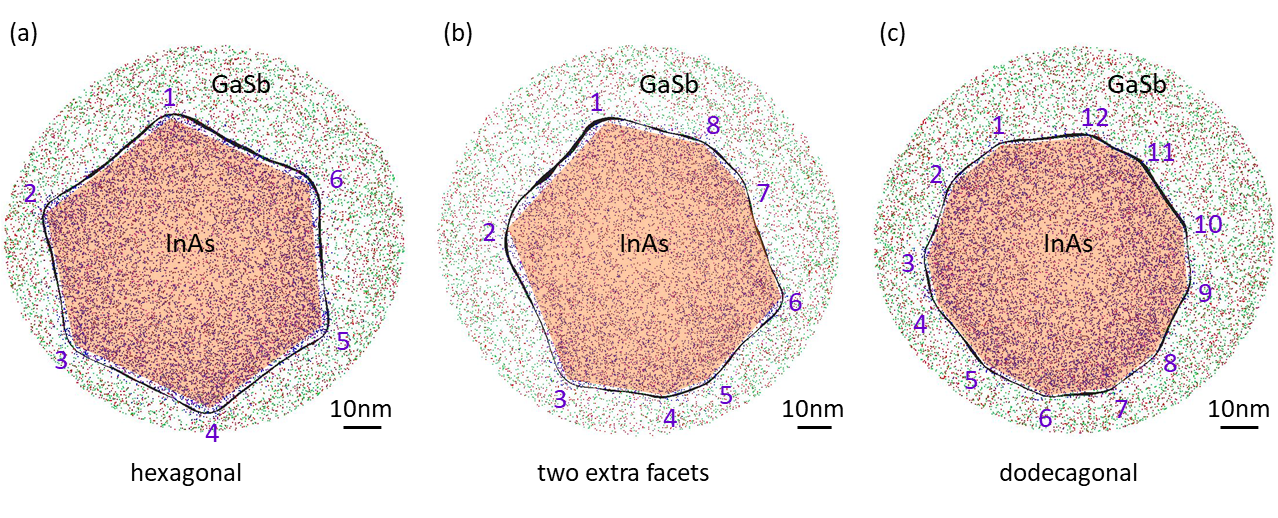}
	\caption[]{Cross section of InAs nanowires with different doping as obtained from 10\,nm slices of APT measurements. The shape of the core is marked using 40\% Indium iso-concentration surfaces\cite{Hellman03} and facets/corners are then marked  manually. (a) Wires with low doping concentration (growth run H) always show hexagonal facets. (b) As the doping increases (growth run I) extra facets start appearing locally. (c) In highly doped samples (growth run J), there are sections with length of hundreds of nanometers which have dodecagonal shape.}
	\label{fig:APT-dodeca}
\end{figure*}

For APT, single InAs nanowires with different Te doping levels and a GaSb shell are isolated using a method described in detail in Ref.~[\onlinecite{Koelling17}]. They are analyzed in a LEAP4000X~HR equipment. We imaged the distribution of Te in the InAs nanowires utilizing the same methods for noise level suppression and background correction as in Ref.~[\onlinecite{Koelling17}]. The APT analyses reveal four notable findings. First, as can be seen in Fig.~\ref{fig:APT-cross}, Te accumulates at the core of the wire and at the corners of the hexagonal \{110\} facets for all investigated Te doping concentrations. 
Second, the measured Te dopant incorporation increases with the corresponding nominal doping (cf. Figs.~\ref{fig:APT-cross} and \ref{fig:APT-conc}). Third, the Te concentration increases towards the bottom of the nanowires typically reaching a factor of $2-3$ higher concentration near the bottom with respect to the top (cf. Fig.~\ref{fig:APT-conc}). Forth, as depicted in Fig.~\ref{fig:APT-dodeca}, with the increase of Te nominal doping, the facets of the hexagonal InAs nanowires become unstable and the nanowires with the highest doping have locally dodecagonal cross sections. This observation is confirmed by TEM cross sectional images, as shown in Supplemental Material Fig.~S1.

We interpret these findings as follows. Te diffuses along the sidewall of the nanowire, a contribution coming also from the substrate surface, and incorporates via the small (111) growth facet on the top of the nanowire (Fig.~\ref{fig:TEM} (b)) and on the corners of the hexagonal facets. Note that the diameters of the top (111) facet in Fig.~\ref{fig:TEM} (b) and the Te rich core in Fig.~\ref{fig:APT-cross} (c) are comparable.  As the wire grows, the diffusion contribution of the substrate surface disappear and Te incorporation at top decreases. Due to the large covalent radius of Te atoms, they accumulate on the corners between facets. The accumulation at the corners promotes  the formation of (112) facets and hence the dodecagonal segments. One can presume that the surfactant Te incorporates easier into the solid on (111) and (112) surfaces than (110). 

\subsection{Electrical Characterization}

\begin{figure}[!ht]
	\centering
\includegraphics[width=1.0\columnwidth]{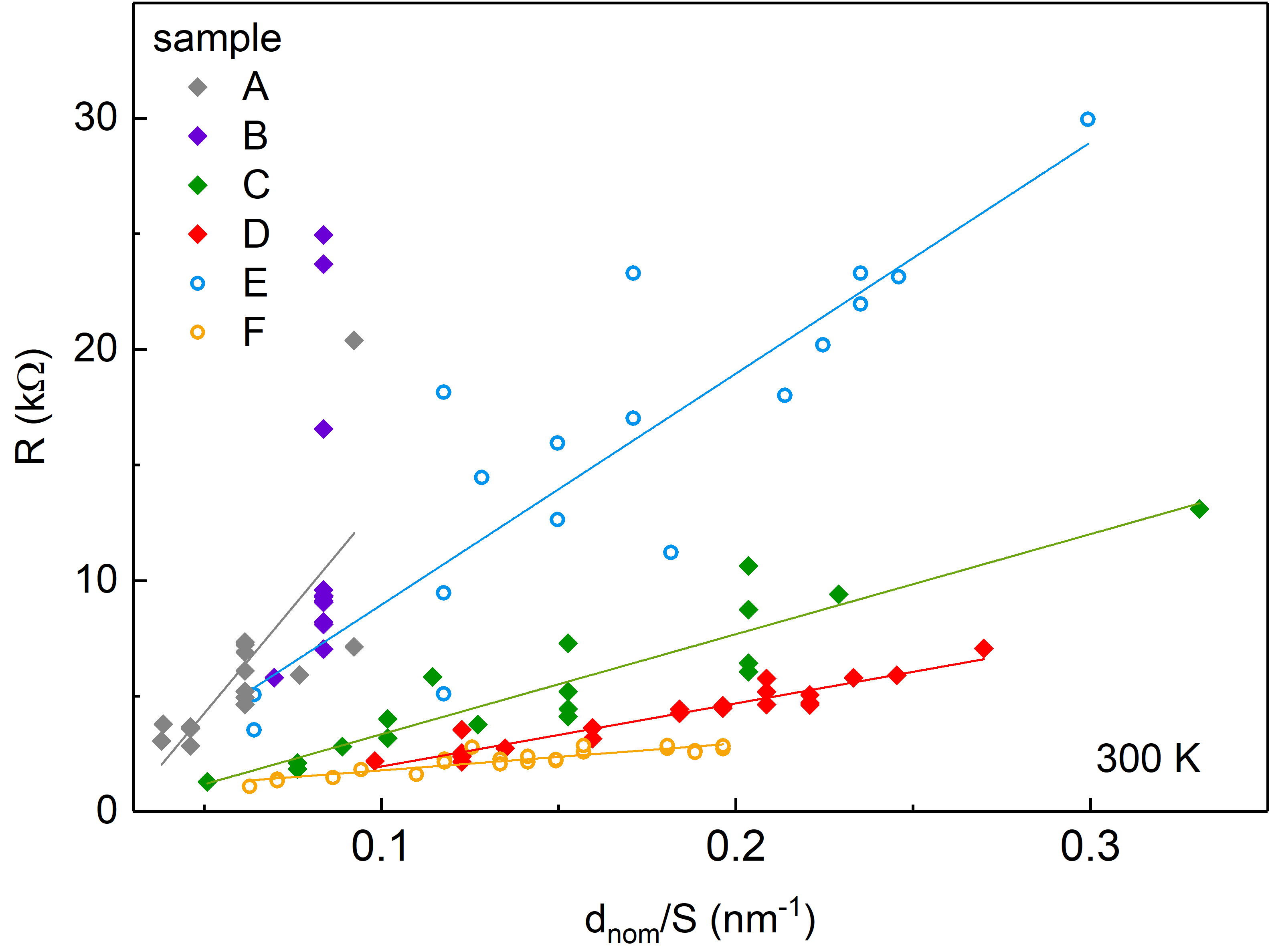}
	\caption[]{Four-terminal resistance $R$ of doped nanowires at room temperature plotted function of the ratio between nominal contact distance $d_\mathrm{nom}$ and mean nanowire cross sectional area $S$. The solid lines represent the linear fits to determine the resistivity. For the samples of growth run B no reliable linear fit was possible.}
	\label{fig:R-4P}
\end{figure}

In Fig.~\ref{fig:R-4P} the resistance at room temperature from nanowires with different doping levels are plotted function of the ratio of nominal distance between the contacts and the cross sectional area $d_\mathrm{nom}/S$. 
Here, $S$ was determined for each growth series by average the diameters of the measured nanowires assuming a regular hexagonal shape. The transport measurements have been carried out in a four-terminal scheme in order to eliminate the effect of contact resistances. For each growth run with a specific doping concentration, one finds a linear increase of resistance with the contact separation length verifying the ohmic behavior of the transport in the wire. The doping effect of Te is confirmed by the decrease of the slope, i.e. decreased resistivity $\rho$, with increasing doping concentration (c.f. Table~\ref{tab:InAs-samples}). In Fig.~\ref{fig:R-4P} one observes that there is a spread of the resistance of different nanowires with identical contact separation. The effect is less pronounced for higher doping levels. One reason for the varying conductance could be found in non-uniform cross section $S$. Although the nanowires were grown in a selective-area epitaxy scheme, fluctuations of the cross sectional area can not be avoided. 

Charge carrier concentrations of nanowires with different doping levels were extracted from field-effect measurements by biasing a global back gate. We found that the conductance is reduced for negative gate voltages, confirming the $n$-type doping character. Quantitative information about charge carrier concentration was obtained using the threshold voltage $V_\mathrm{th}$ at pinch-off. Since nanowires with a Te doping concentration larger than $5 \times 10^{18}\,\mathrm{cm}^{-3}$ did not reach pinch-off at accessible back-gate voltages, $V_\mathrm{th}$ was in this case obtained by extrapolating the approximately linear dependence towards positive voltages. Example of traces including a linear fit are given in the Supplemental Material in Figs.~S2 - S4. Some nanowire samples showed a gate hysteresis behavior, which was reproducible during several cycles. In that case, $V_\mathrm{th}$ was determined by taking the average of the threshold voltages for up- and down-sweeps. Details of the determination of $V_\mathrm{th}$ and the calculation of the charge carrier concentration $n_\mathrm{3D}$ are given in the Supplemental Material.
\begin{figure}[!ht]
\includegraphics[width=1.0\columnwidth]{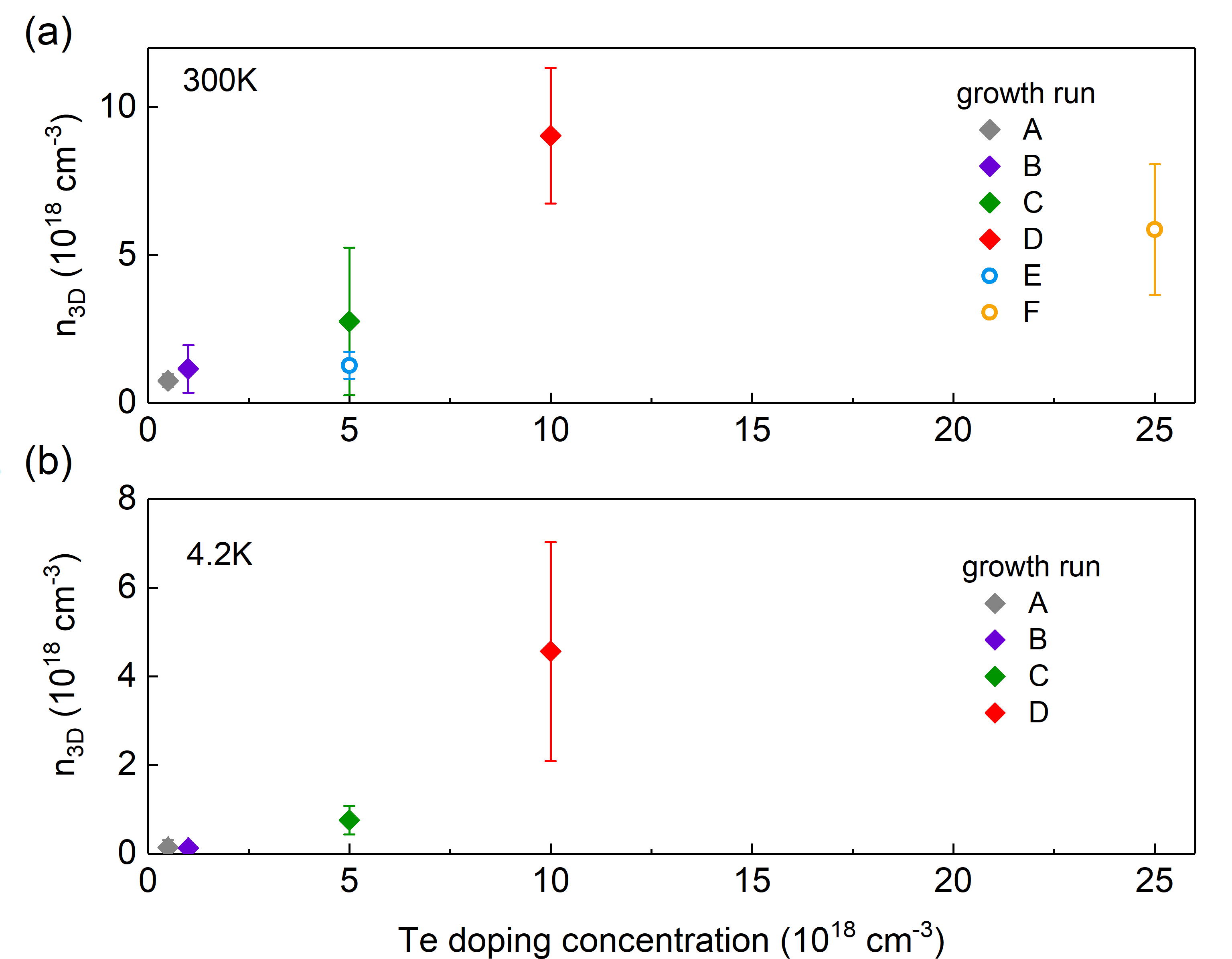}
	\caption[]{(a) Extracted charge carrier concentrations $n_\mathrm{3D}$ from gate pinch-off curves at room temperature for nanowires with different nominal Te-doping concentrations (growth runs A to F). (b) $n_\mathrm{3D}$ at 4.2\,K for growth runs A to D.}
	\label{fig:n3d-300K-4K}
\end{figure}

The field-effect measurements on nanowires grown with different Te doping concentrations were carried out at room temperature and at 4.2\,K. The corresponding charge carrier concentrations at different nominal doping concentrations are depicted in Fig.~\ref{fig:n3d-300K-4K}. The graphs confirm that the experimentally determined charge carrier concentration values increase with the nominal doping of the nanowires. 

For sample C and D, one finds that at room temperature, the measured carrier concentrations are close to the nominal Te doping concentrations demonstrating the efficient dopant incorporation. However, for samples E and F grown with the low As$_4$ flux the doping efficiency is somewhat lower, the obtained charge carrier concentrations being lower than expected from the previous growth runs A to D. This could be explained by a modified Te incorporation due to altered growth conditions by decreased As$_4$ flux. The measured charge carrier concentrations are found to be generally higher at room temperature compared with 4.2\,K, which represents  the expected behavior in semiconductors.

A relatively large spread of $n_\mathrm{3D}$ for the investigated nanowires from identical batches is observed. A possible explanation for that is the imprecise character of the method to determine the charge carrier concentration from the gate pinch-off threshold voltage. Defects within the oxide and at the interface with the semiconductor effectively alter the capacitance used to determine the carrier concentration (see section II-B in the Supplemental Material)  leading to scattered results. Furthermore, it was shown in Fig.~\ref{fig:APT-conc} that the Te incorporation along the nanowire axis is not uniform, resulting in different effective doping depending on the length of the particular nanowire. Therefore, one must be careful to directly compare the obtained results with measurements from APT. However, it can be concluded that the charge carrier concentration strongly depends on the nominal dopant concentration, suggesting an effective $n$-type doping of InAs by Te.

\subsection{Josephson junctions}

\begin{figure*}[!ht]
	\centering
\includegraphics[width=0.90\textwidth]{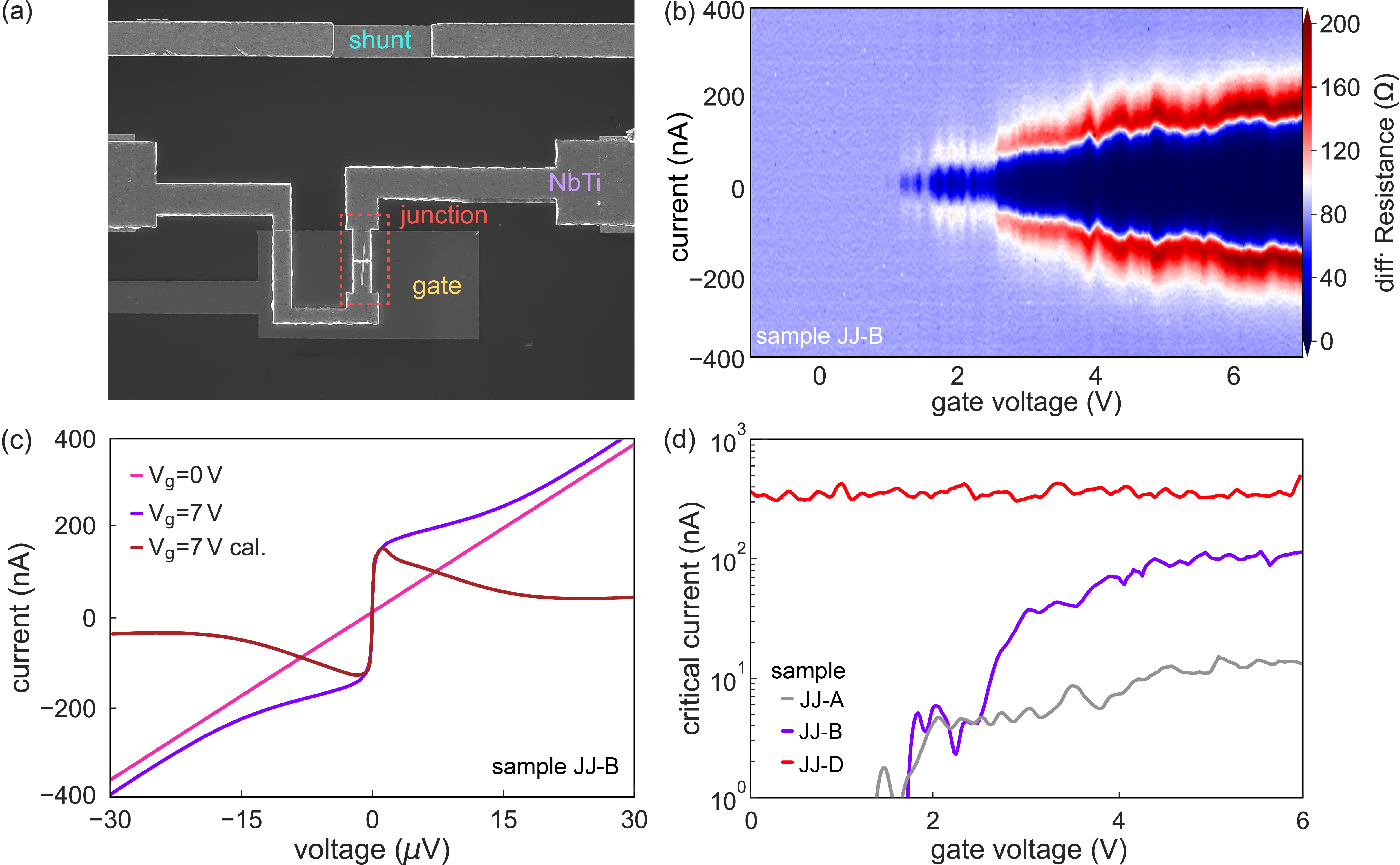}
	\caption[]{(a) SEM micrograph of a nanowire Josephson junction that is connected in parallel to an extrinsic on-chip shunt resistor. The superconducting leads consist of a 80-nm-thick NbTi layer. The electrostatic tuning is realized by means of a global bottom gate pad. (b) Gate- and current-dependent differential resistance map of a nanowire Josephson junction of growth run B. The measurements were taken at 15\,mK. (c) Single voltage-current characteristics taken from (a) at gate voltages of $V_\mathrm{g}=0\,$V and 7\,V. The calibrated characteristics for $V_\mathrm{g}=7\,$V after subtracting the conductance of the shunt is also given. (d) Gate-dependent critical current for junctions JJ-A, JJ-B, and JJ-D at 15\,mK.}
	\label{fig:Junctions_Gate}
\end{figure*}

In order to demonstrate the suitability of Te-doped nanowires for superconducting hybrid structures the properties of gate-controlled Josephson junctions are investigated. The measurements have been performed in a $^3$He/$^4$He dilution refrigerator with a base temperature of 15\,mK. The measured junctions JJ-A, JJ-B, and JJ-D were fabricated from nanowires from growth series A, B, and D, respectively. A typical structure is shown in Fig.~\ref{fig:Junctions_Gate} (a),
in which the nanowire-based Josephson junction is shunted by an AuGe resistor. 
The shunt was included to suppress hysteretic effects in the current-voltage characteristics and to improve the performance of the device in measurements of the ac-Josephson effect \cite{Joyez99,Chauvin05}. In Fig.~\ref{fig:Junctions_Gate} (b) the color-coded differential resistance vs. bias current is plotted in the gate voltage range from $V_\mathrm{g}=-1$\,V to 7\,V for junction JJ-B made from a nanowire having a doping nominal concentration of $1 \times 10^{18}$\,cm$^{-3}$. One finds that for gate voltages below $1\,$V the nanowire is pinched off, with the remaining resistance of 80\,$\Omega$ given by the shunt resistor. Between $V_\mathrm{g}$=$1\,$V and $2.5\,$V the device shows signatures of Coulomb blockade which can be attributed to the presence of an intrinsic quantum dot.\cite{Zellekens20a} At gate voltages above $2.5\,$V, the nanowire channel is open and exhibits gate-tunable supercurrent, i.e., an important requirement for hybrid superconducting circuit applications. The critical current shows some superimposed fluctuations due to interference effects.\cite{Doh05,Guenel12} From Fig.~\ref{fig:Junctions_Gate} (c) showing a single voltage-current ($I-V$) trace at $V_\mathrm{g}=7$\,V (red curve) a critical current of $I_\mathrm{c}=140\,$nA is extracted. The $I-V$ characteristics contains the contribution of the shunt resistor, which needs to be subtracted to gain the actual current through the junction. Indeed, the resistance of the shunt can be determined directly from the measurement at zero gate voltage (cf. Fig.~\ref{fig:Junctions_Gate} (c)), where the junction is pinched-off completely. After the calibration of the junction response by means of a point-wise combination of the traces at $7$\,V and $0$\,V, it is possible to extract the unperturbed characteristics of the device (cf. Fig.~\ref{fig:Junctions_Gate} (c), dark red trace). Junctions JJ-A and JJ-B could be pinched-off completely so that the normal state resistance $R_\mathrm{N}$ could be extracted. At a gate bias of $7$\,V we determined $I_\mathrm{c}R_\mathrm{N}$ products of 65\,$\mu$V and $188\,\mu$V for sample JJ-A and JJ-B, respectively.

The critical current as a function of gate voltage for all investigated Josephson junctions are shown in Fig.~\ref{fig:Junctions_Gate} (d). The junction containing a nanowire with the highest doping level only shows a weak relation between $V_\mathrm{g}$ and the measured $I_\mathrm{c}$. In contrast, the wires with the lower carrier concentrations exhibit a transistor-like behavior with pronounced pinch-off and saturation regions. All critical currents shown some fluctuations due to inference effects in the nanowire channel.\cite{Doh05,Guenel12} The difference in the threshold voltage between junction JJ-A and JJ-B are attributed to Coulomb resonances which dominate the transport at low gate values.\cite{Zellekens20a} A comparison of the critical current at $V_\mathrm{g}=6\,$V reveals increase by a factor of 10 between JJ-A and JJ-B and a factor of five between JJ-B and JJ-D, which confirms the impact of the doping concentration on $I_\mathrm{c}$.

\begin{figure*}[!ht]
	\centering
\includegraphics[width=0.99\textwidth]{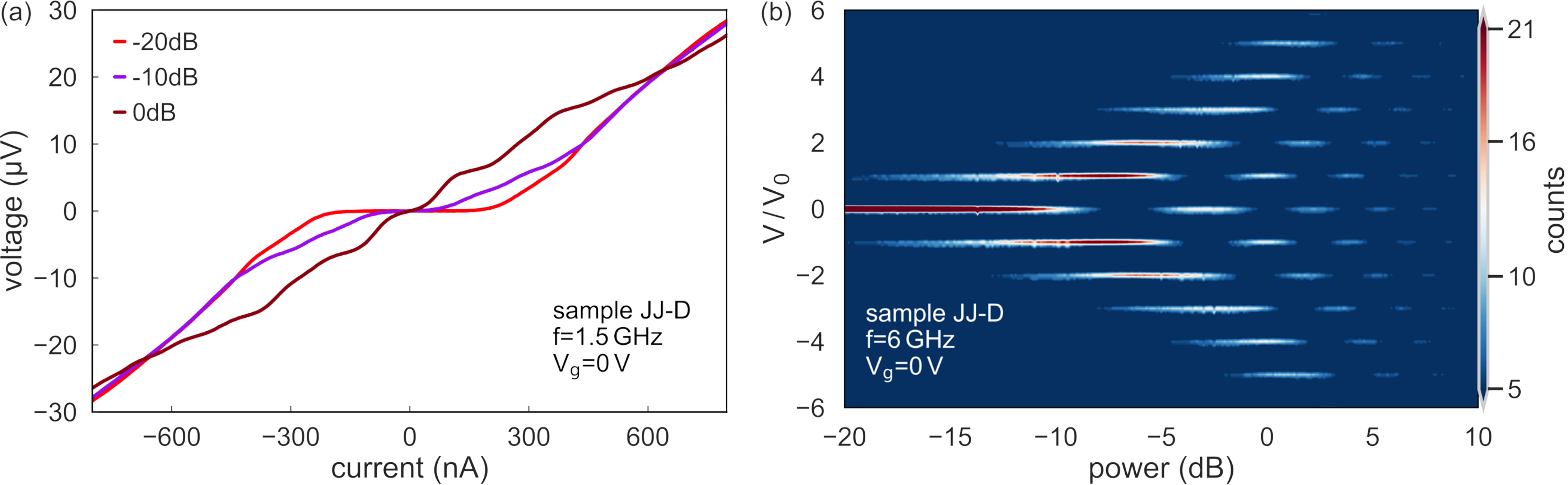}
	\caption[]{(a) Current-voltage characteristics of junction JJ-D for different microwave powers. (b) Measurements of Shapiro steps as a function of power. The color scale corresponds to the number of measurement points at normalized voltages $V/V_0$ with $V_0=hf/(2e)$.}
	\label{fig:Shunted_junctions:Shapiro_spectra}
\end{figure*}

A reliable way to confirm that nanowire-based junction do indeed carry a Josephson supercurrent is to perform measurements under microwave irradiation. Here, the application of a microwave signal results in Shapiro steps of height $n \cdot V_0$, with
$V_0=hf/(2e)$, $h$ Planck constant and $n = 1, 2, 3, \dots$. Figure~\ref{fig:Shunted_junctions:Shapiro_spectra} (a) shows a set of current–voltage traces of junction JJ-D for $f = 5$\,GHz and different microwave powers at $V_\mathrm{g}=7$\,V. 
At low power, i.e. $-20$\,dB, the curve basically mimics the behavior of a purely DC-driven junction. However, if the power is increased to $-10$ \,dB and further to $0$\,dB, the zero voltage state is gradually suppressed and equidistant voltage plateaus, i.e. Shapiro steps, appear. In Fig.~\ref{fig:Shunted_junctions:Shapiro_spectra} (b) the presence of Shapiro steps at multiples of $V_0$ for $f=6$\,GHz is shown as a function of microwave power. The color scale gives the number of counts as a function of voltage, i.e. large number of count corresponds to the appearance of a step. The regular oscillating pattern without any subharmonic features indicates a sinusoidal current-phase relation of the Josephson junction.

\section{Conclusions}

We demonstrate that Te is an efficient dopant for MBE-grown InAs nanowires. At very high dopant concentrations, above $1 \times 10^{19}$\,cm$^{-3}$, a change in wire morphology was observed, where the hexagonal cross section changes to a dodecagonal one. The side facets have  also a great influence on the distribution of Te dopant, which accumulates at the corners of the hexagonal (111) facets,as APT revealed. In addition, the APT showed that in the center of the wire the Te concentration is higher and that the total Te concentration increases toward the bottom of the nanowire. The effective doping was confirmed by electrical measurements at room temperature and 4\,K, where a systematic increase in the conductivity of the wire with the doping concentration was observed. Indeed, a significant increase in the critical current was obtained in Josephson junctions with a nanowire weak link. At moderate doping concentrations, even gate control was maintained. Since InAs nanowires are often used in hybrid structures for Majorana physics or quantum computing, Te doping provides a very efficient method to tailor the nanowire properties for devices in these applications.

\section*{Acknowledgements}

We thank Tobias Ziegler, Patrick Liebisch, Patrick Pilch, Bram Aarts and Abbas Espiari for their assistance during sample preparation and Michael Schleenvoigt for the metal deposition, Marvin Jansen for initial assistance with TEM, Benjamin Bennemann, Christoph Krause, and Herbert Kertz for technical assistance. Dr. Florian Lentz and Dr. Stefan Trellenkamp are gratefully acknowledged for electron beam lithography. Dr. Elmar Neumann and Stephany Bunte for their assistance with the focused ion beam setup and scanning electron microscopy. Dr. Gianluigi Catelani is gratefully acknowledged for theory support regarding the magnetic field measurements. All samples have been prepared at the Helmholtz Nano Facility.\cite{HNF17} The work at Forschungszentrum J\"ulich was partly supported by the project "Scalable solid state quantum computing", financed by the Initiative and Networking Fund of the Helmholtz Association, by the Deutsche Forschungsgemeinschaft (DFG, German Research Foundation) Project: "Coupling of quantum dots with superconductors", (SCHA 835/9-1), UK EPSRC is acknowledged for funding through grant No. EP/P000916/1. The APT work was supported by NSERC Canada (Discovery, SPG, and CRD Grants), Canada Research Chairs, Canada Foundation for Innovation, Mitacs, PRIMA Québec, and Defence Canada (Innovation for Defence Excellence and Security, IDEaS). The work at RIKEN was partially supported by Grant-in-Aid for Scientific Research (B) (No. 19H02548), Grants-in-Aid for Scientific Research (S) (No. 19H05610), and Scientific Research on Innovative Areas "Science of hybrid quantum systems" (No. 15H05867). 

\clearpage

\pagebreak 


\setcounter{section}{0}
\setcounter{equation}{0}
\setcounter{figure}{0}
\setcounter{table}{0}
\setcounter{page}{1}
\makeatletter
\renewcommand{\thesection}{S\Roman{section}}
\renewcommand{\thesubsection}{\Alph{subsection}}
\renewcommand{\theequation}{S\arabic{equation}}
\renewcommand{\thefigure}{S\arabic{figure}}
\renewcommand{\figurename}{Supplementary Figure}
\renewcommand{\bibnumfmt}[1]{[S#1]}
\renewcommand{\citenumfont}[1]{S#1}

\begin{center}
\textbf{Te-doped selective-area grown InAs nanowires for superconducting hybrid devices (Supplemental Material)}
\end{center}

\section{TEM cross-section images showing nanowires facets}

\begin{figure}[!ht]
	\centering
\includegraphics[width=0.95\columnwidth]{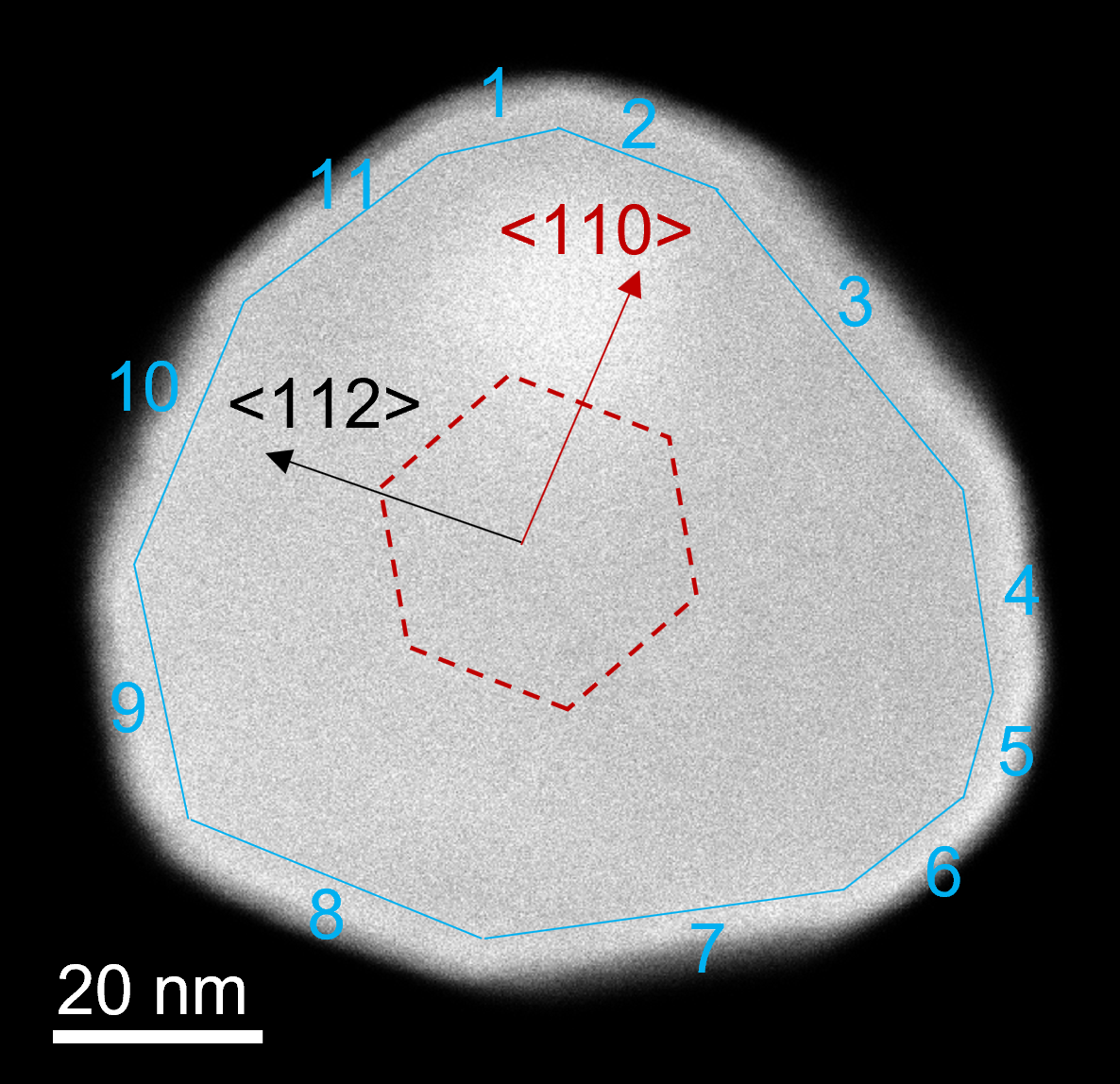}
	\caption{A cross section taken from the middle of a nanowire core with the highest doping concentration ($2.5 \times 10^{19}\,\mathrm{cm}^{-3}$) of growth run F, showing formation of multiple facets of $\{110\}$ and $\{112\}$ (11 in this case) leading to the dodecagonal shape.}
	\label{fig:TEM-Te-doping-SI}
\end{figure}

In Fig.~\ref{fig:TEM-Te-doping-SI} a cross section of a nanowire from growth run F is depicted. 

\section{Electrical Characterization}

\subsection{Electrical characterization of Te-doped nanowires}

Electrical characterisation of nanowires with varying Te-doping levels were performed at room temperature as well as at 4.2\,K. Room temperature measurements without back gate were done using a Keithley SCS-4200 semiconductor characterisation system connected to a 4-point probe station. Measurements at low temperature were performed in a Lakeshore liquid $^4$He flow-cryostat capable of cooling down to approximately 4.0\,K. Nanowires were investigated in two- and four-terminal configurations with and without applying a voltage to the back gate. A Keithley 2636B source-measurement unit was used for the application of the bias current. Potential differences in four-terminal configuration were measured using a Keysight 34461A multimeter with input impedance of 10\,M$\Omega$.

\subsection{Determination of carrier concentration}

\begin{figure}[!ht]
\centering
\includegraphics[width=0.9\columnwidth]{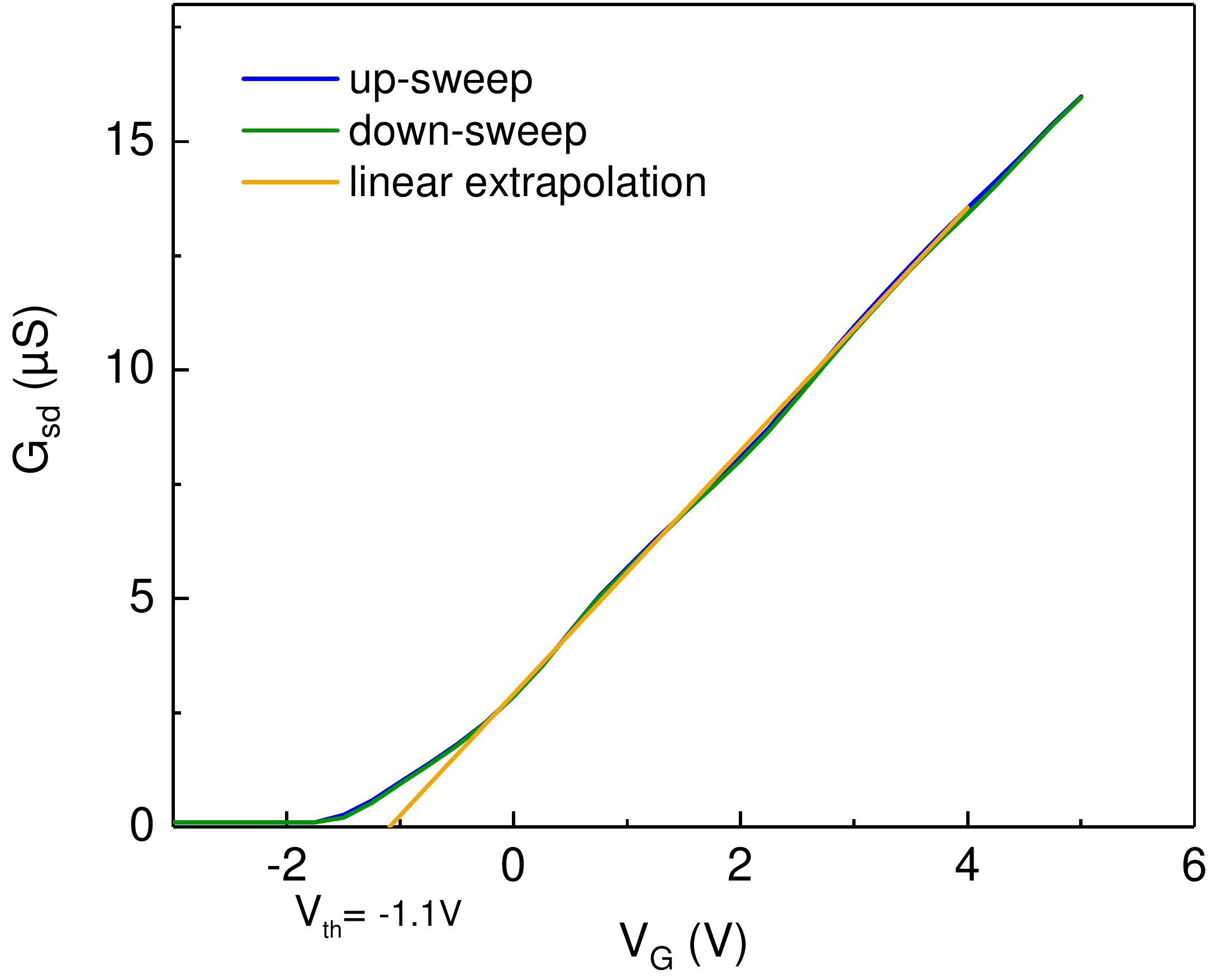}
\caption[]{Representative source-drain conductance $G_{sd}$ of a nanowire with nominal doping $1\times 10^{18}\,\mathrm{cm}^{-3}$ (growth run B) depending on back-gate voltage. Threshold voltage $V_\mathrm{th}$ is extrapolated from the approximately linear behaviour at positive gate voltages.}
	\label{fig:G_4P}
\end{figure}

\begin{figure}[!ht]
\centering
\includegraphics[width=0.9\columnwidth]{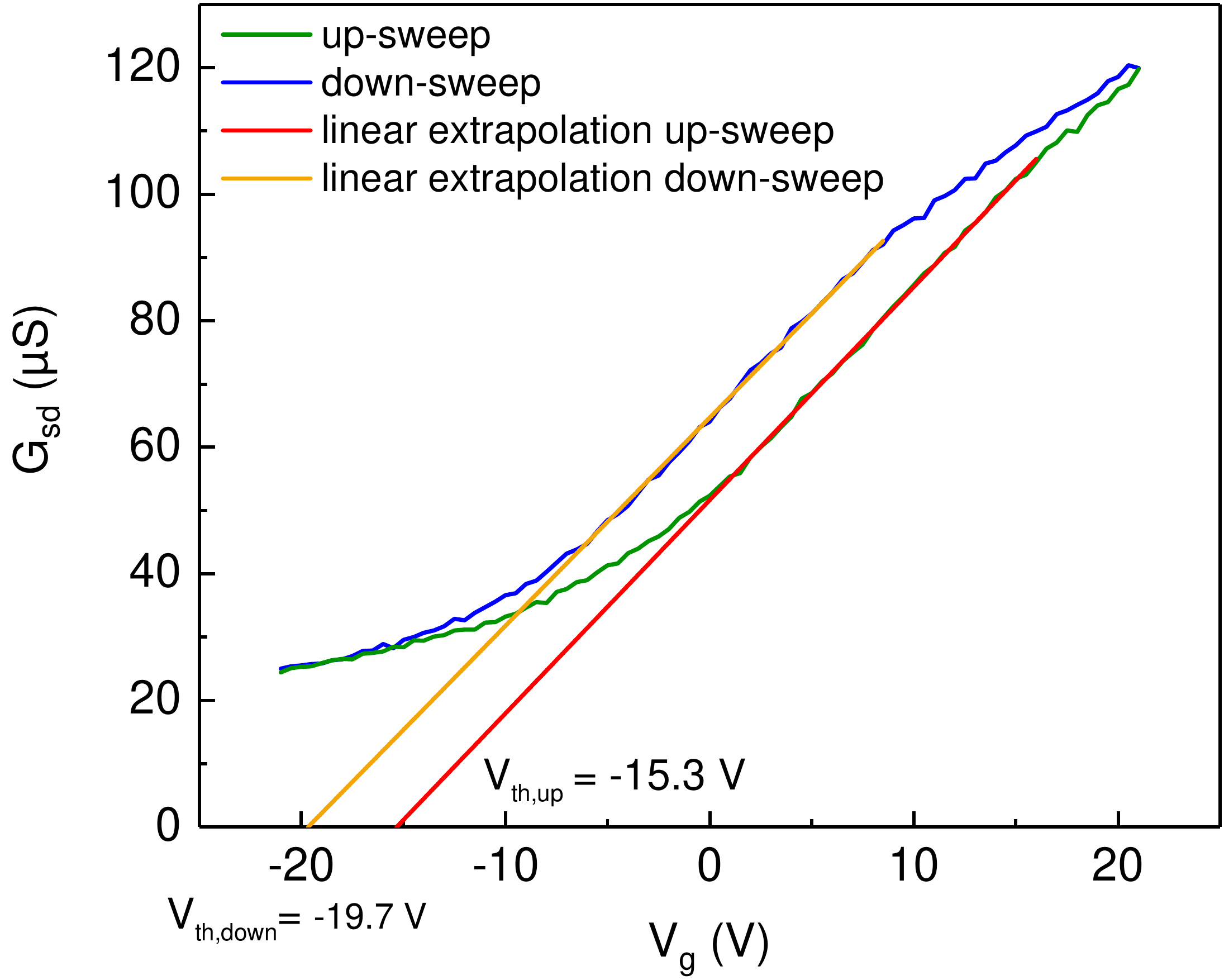}
\caption[]{Representative source-drain conductance $G_{sd}$ of a nanowire with nominal doping $5\times 10^{18}\,\mathrm{cm}^{-3}$ (growth run E) depending on back-gate voltage $V_\mathrm{g}$. The threshold voltage $V_\mathrm{th}$ is extrapolated from the approximately linear behaviour at positive gate voltages. The observed conductance shows a hysteresis effect in between gate up- and down-sweep, resulting in different obtained threshold voltages. For further analysis the mean value from up- and down-sweep was used.}
	\label{fig:G_4P_Hysterese}
\end{figure}

\begin{figure}[!ht]
	\centering
\includegraphics[width=0.9\columnwidth]{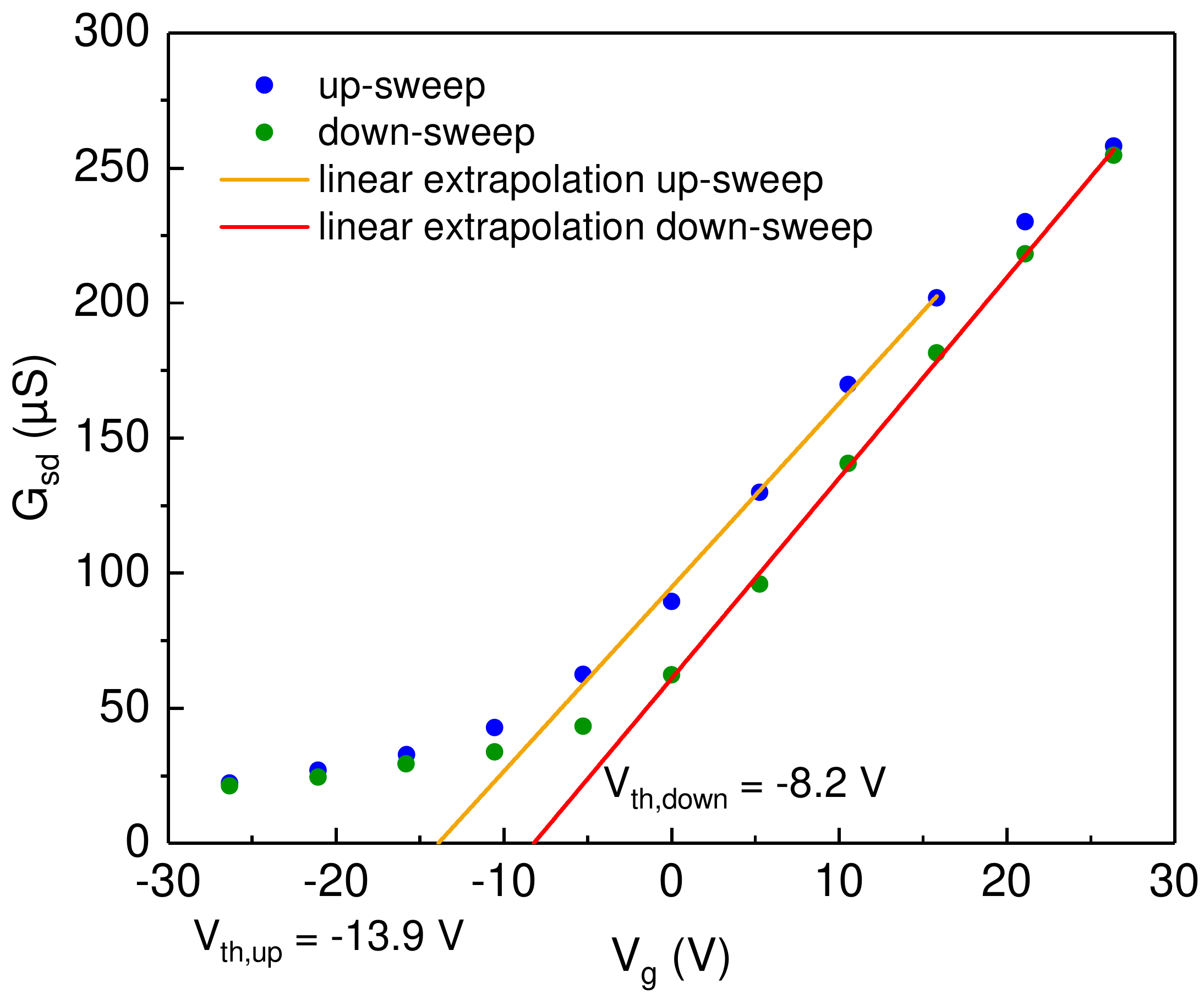}
\caption[]{Representative source-drain conductance $G_{sd}$ for a nanowire with nominal doping $5\times 10^{18}\,\mathrm{cm}^{-3}$ (growth run C), extracted from $IV$ measurements at varying back-gate voltages for both gate up- and down sweep. Threshold voltages V$_{th}$ are extrapolated from the approximately linear behaviour at positive gate voltages.}
	\label{fig:G_4P_IV}
\end{figure}

In order to obtain information on charge carrier density in the nanowires, the source-drain conductance was investigated depending on voltage $V_\mathrm{g}$ applied to the global back-gate electrode. The threshold voltage $V_\mathrm{th}$ was extracted by extrapolating from the linear behavior at positive gate voltages (cf. Fig.~\ref{fig:G_4P}). Some gate-dependent measurements showed an hysteresis effect, i.e. the results for gate up- and down-sweep were slightly shifted. In these cases, the average value was used for further evaluation (c.f. Figs.~\ref{fig:G_4P_Hysterese} and \ref{fig:G_4P_IV}). Gate-dependent measurements were achieved on two ways. On the one hand, the gate was swept by directly taking the source-drain current at constant $V_{sd}$. Alternatively, the conductance was extracted from $IV$ sweeps at different fixed gate voltages. This was preferred in cases with a small gate hysteresis, in order to reduce uncertainties in conductance from offsets of the measurement devices. In Fig.~\ref{fig:G_4P_IV} an example measurement for the latter case is shown.

For calculating the charge carrier concentration from the threshold voltages the following relation was used\cite{Dayeh07}
\begin{align}
    n_\text{3d} = \frac{C\left|V_\mathrm{th}\right|}{el_\mathrm{NW}\pi r_\mathrm{NW}^2} \, , 
\label{eq:N3D}
\end{align}
with $e$ the electron charge, $l_\mathrm{NW}$ the nanowire length  between the contacts, $r_\mathrm{NW}$ its radius. $C$ denotes the capacitance between the nanowire and the global back-gate electrode. It is given by\cite{Dayeh07}
\begin{equation}
    C = \frac{2\pi\epsilon_0\epsilon l_\mathrm{NW}}
    {\ln \left[ \left(2h+d+2\sqrt{h^2+hd}\right)/d\right]} \, . 
\label{eq:C_N3D}
\end{equation}
Here, $\epsilon_0$ is the dielectric constant, $\epsilon$ is the relative dielectricity of the gate oxide, i.e. $\epsilon = 3.9$ for SiO$_2$, $h$ denotes the thickness of the oxide between back-gate and nanowire and $d = 2r_\text{NW}$ the diameter of the nanowire.

\section{Extrinsically shunted Josephson junctions}

\begin{figure*}[!t]
	\centering
\includegraphics[width=0.85\textwidth]{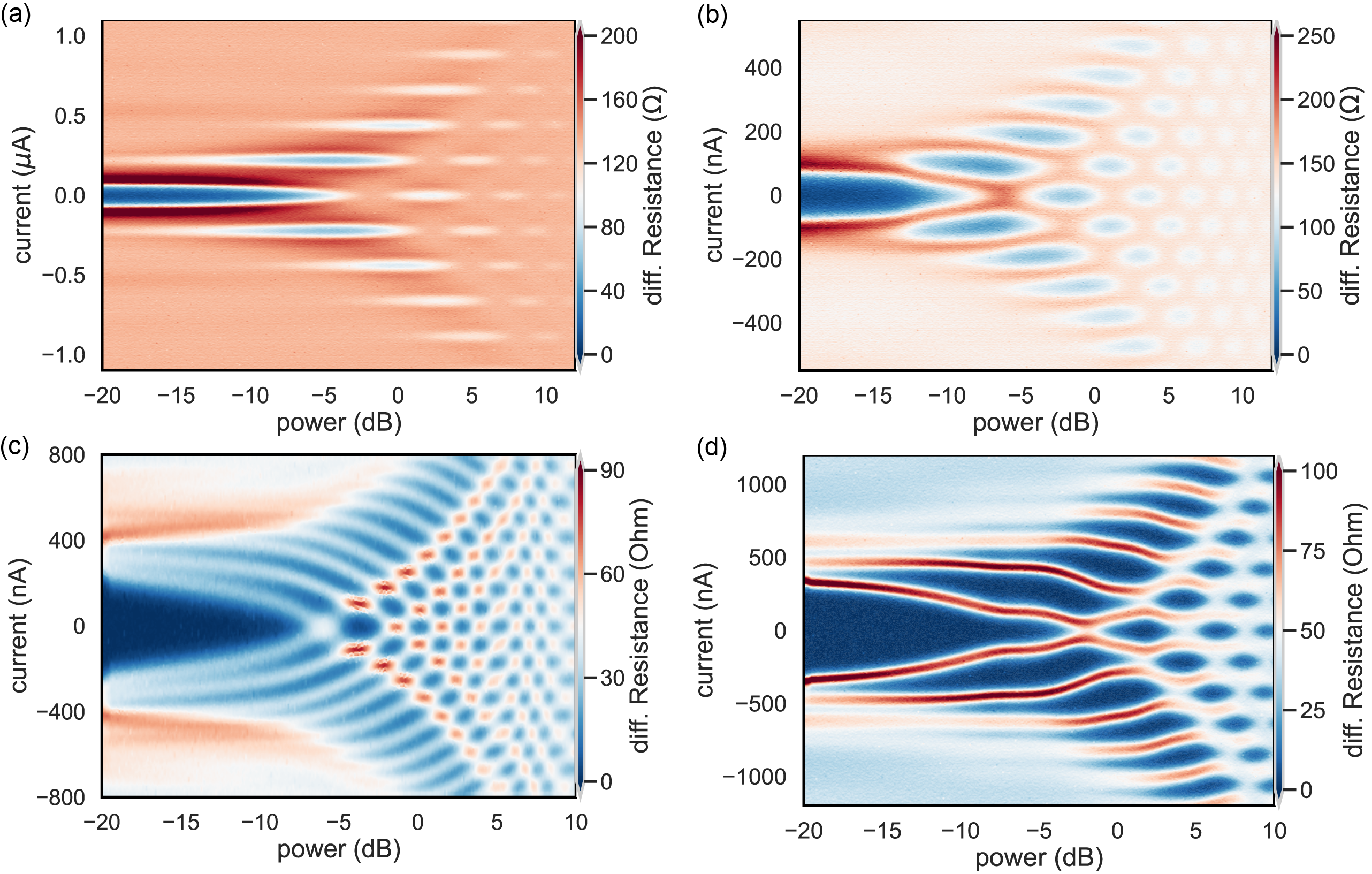}
	\caption[]{Shapiro response of samples JJ-A  at  (a) f$\,$=$\,$4.1$\,$GHz and (b) f$\,$=$\,$6.0$\,$GHz and JJ-D at  (c) f$\,$=$\,$1.5$\,$GHz and (d) f$\,$=$\,$3.8$\,$GHz for a gate voltage of V$_\mathrm{g}$=7$\,$V [(a),(b)] and V$_\mathrm{g}$=0$\,$V [(c),(d)].  } 
	\label{fig:Shunted_junctions:Shunt_calibration_combined}
\end{figure*}

In Josephson junctions the dissipationless supercurrent arises from the coherent tunneling of Cooper pairs between two superconducting electrodes, driven by the phase difference $\delta$. In tunnel junctions with large spatial extensions, the phase can be considered as a quasistatic parameter due to the damping effect of the large capacitance associated with it. The large time constant suppresses sudden changes of the voltage $V\propto \partial\delta/\partial t$, which stabilizes the system with respect to thermal and quantum fluctuations. However,if one assumes a nanowire Josephson junction, the phase dynamics are not longer purely determined by the superconductor/nanowire/superconductor stack itself. Instead, the superconducting leads and the circuit environment, modeled as a admittance $Y(\omega)$, with $\omega$ the frequency, start to influence the electromagnetic properties of the device. Thus, the phase difference $\delta$ between the two superconducting electrodes of the junction is not a simple parameter anymore rather than the combination of all phase fluctuation contributions\cite{Joyez99,Chauvin05}
\begin{equation}
\sqrt{\langle\delta^2\rangle}\propto \frac{Z_s(\omega )}{R_\mathrm{K}\left( 1+ e^{-\hbar \omega /k_B T}\right)} \, . 
\end{equation}
Here, $R_K$ corresponds to the von-Klitzing constant ($h/e^2$), $T$ is the electron temperature and $Z_s(\omega)$ is the total impedance. The latter is thereby determined by the junction inductance $L_0$, the junction capacitance $C_0$ and the impedance of the external circuit. In case of a small junction, $L_0$ and $C_0$ are small, too, and $Z_s(\omega)$ is dominated by the external circuit admittance $Y(\omega )$. Such a scenario is usually associated with large phase fluctuations, which ultimately lead to a stochastic behavior of the corresponding supercurrent:
\begin{equation}
I(\omega)=\langle I_0\rangle = I_0 \langle \sin{\left( \delta \right)}\rangle \, , 
\end{equation}
e.g. a widely spread distribution of the switching current or the occurrence of a hysteresis in the zero-voltage state. The latter is especially important in the case of large supercurrents, for which the actual electron temperature deviates significantly from the bath temperature due to overheating effects.\cite{Courtois08}

The most simple way to limit both phase noise as well as phase diffusion is the implementation of an on-chip shunt resistor in close proximity to the nanowire Josephson junctions.\cite{Joyez99,Chauvin05} In our case, the InAs/Al half-shell nanowire Josephson junction is shunted by an on-chip AuGe stripe (cf. Fig.~8 (a) in the main text) Both elements are connected by superconducting TiN electrodes. The normal conducting shunt resistor causes as suppression of the pronounced and abrupt switch between zero voltage state and dissipative transport. Instead, there is a smooth and continuous transition between both branches, effectively providing experimental access to the full voltage range. After the subtraction of the shunt resistance, the remaining trace exhibits the typical decrease of the current above a critical value, which is a clear signature for the suppression of the Cooper pair driven transport. By extracting the turning point of the supercurrent peak, as shown in, it is then possible to obtain the maximum junction current $I_0$.\cite{Joyez99,Chauvin05} 

The full calibration procedure is as follows: 
First, the gate dependency of the supercurrent is investigated in order to measure the shunt resistor independently. Thus, the nanowire has to be pushed into pinch-off by applying a sufficiently large negative gate voltage. Now, by applying a linear fit within the range of the supercurrent branch to every individual $I$-$V$ characteristics, it is possible to obtain the combined resistance of the parallel circuit. For negative voltages, i.e. when the nanowire is pinched-off, the circuit resistance becomes equal to the normal conducting shunt resistance. In order to limit the influence of noise and measurement uncertainties, the whole saturation region is used to calculate an average value for the resistor. The contribution of the normal conducting shunt resistor, which acts as an additive and linear contribution, can then be removed by a point-wise extraction of the junction resistance $R_\mathrm{JJ}$ and calibration of the measured voltage $V$
\begin{eqnarray}
R_\mathrm{JJ}&=&\frac{V R_\mathrm{s}}{I_\mathrm{bias}R_\mathrm{s}-V_{}} \, , \\
I_\mathrm{JJ}&=&\frac{V_{}}{R_\mathrm{JJ}} \, , 
\end{eqnarray}
with $I_\mathrm{bias}$ as the externally applied current. \\

\section{Shapiro measurements on nanowire Josephson junctions with different doping concentrations}

Figure \ref{fig:Shunted_junctions:Shunt_calibration_combined} provides an overview of the power-dependent Shapiro response for the junctions JJ-A and JJ-D in the main manuscript. For the undoped nanowires, a pronounced spectrum can be observed for both measurements, i.e. f$\,$=$\,$4.1$\,$GHz (cf. Fig. \ref{fig:Shunted_junctions:Shunt_calibration_combined} (a)) and f$\,$=$\,$6.0$\,$GHz (cf. Fig. \ref{fig:Shunted_junctions:Shunt_calibration_combined} (b)), respectively. Thus, even though the nanowire have a low carrier concentration and despite the comparably poor quality of the ex-situ contacts, a clear supercurrent without any resistive contribution can be observed if a sufficiently large gate voltage (V$_\mathrm{g}>$4$\,$V) is applied.  The same holds for the wires with the highest doping level in Figs. \ref{fig:Shunted_junctions:Shunt_calibration_combined} (c) and (d), respectively, proving that the large supercurrent of I$_\mathrm{s}>$200$\,$nA is really carried by a Josephson junction rather than just a metallic short.


\begin{thebibliography}{45}%
\makeatletter
\providecommand \@ifxundefined [1]{%
 \@ifx{#1\undefined}
}%
\providecommand \@ifnum [1]{%
 \ifnum #1\expandafter \@firstoftwo
 \else \expandafter \@secondoftwo
 \fi
}%
\providecommand \@ifx [1]{%
 \ifx #1\expandafter \@firstoftwo
 \else \expandafter \@secondoftwo
 \fi
}%
\providecommand \natexlab [1]{#1}%
\providecommand \enquote  [1]{``#1''}%
\providecommand \bibnamefont  [1]{#1}%
\providecommand \bibfnamefont [1]{#1}%
\providecommand \citenamefont [1]{#1}%
\providecommand \href@noop [0]{\@secondoftwo}%
\providecommand \href [0]{\begingroup \@sanitize@url \@href}%
\providecommand \@href[1]{\@@startlink{#1}\@@href}%
\providecommand \@@href[1]{\endgroup#1\@@endlink}%
\providecommand \@sanitize@url [0]{\catcode `\\12\catcode `\$12\catcode
  `\&12\catcode `\#12\catcode `\^12\catcode `\_12\catcode `\%12\relax}%
\providecommand \@@startlink[1]{}%
\providecommand \@@endlink[0]{}%
\providecommand \url  [0]{\begingroup\@sanitize@url \@url }%
\providecommand \@url [1]{\endgroup\@href {#1}{\urlprefix }}%
\providecommand \urlprefix  [0]{URL }%
\providecommand \Eprint [0]{\href }%
\providecommand \doibase [0]{http://dx.doi.org/}%
\providecommand \selectlanguage [0]{\@gobble}%
\providecommand \bibinfo  [0]{\@secondoftwo}%
\providecommand \bibfield  [0]{\@secondoftwo}%
\providecommand \translation [1]{[#1]}%
\providecommand \BibitemOpen [0]{}%
\providecommand \bibitemStop [0]{}%
\providecommand \bibitemNoStop [0]{.\EOS\space}%
\providecommand \EOS [0]{\spacefactor3000\relax}%
\providecommand \BibitemShut  [1]{\csname bibitem#1\endcsname}%
\let\auto@bib@innerbib\@empty
\bibitem [{\citenamefont {Dayeh}\ \emph {et~al.}(2007)\citenamefont {Dayeh},
  \citenamefont {Aplin}, \citenamefont {Zhou}, \citenamefont {Yu},
  \citenamefont {Yu},\ and\ \citenamefont {Wang}}]{Dayeh07}%
  \BibitemOpen
  \bibfield  {author} {\bibinfo {author} {\bibfnamefont {S.}~\bibnamefont
  {Dayeh}}, \bibinfo {author} {\bibfnamefont {D.~P.}\ \bibnamefont {Aplin}},
  \bibinfo {author} {\bibfnamefont {X.}~\bibnamefont {Zhou}}, \bibinfo {author}
  {\bibfnamefont {P.~K.}\ \bibnamefont {Yu}}, \bibinfo {author} {\bibfnamefont
  {E.}~\bibnamefont {Yu}}, \ and\ \bibinfo {author} {\bibfnamefont
  {D.}~\bibnamefont {Wang}},\ }\href {\doibase 10.1002/smll.200600379}
  {\bibfield  {journal} {\bibinfo  {journal} {Small}\ }\textbf {\bibinfo
  {volume} {3}},\ \bibinfo {pages} {326} (\bibinfo {year} {2007})}\BibitemShut
  {NoStop}%
\bibitem [{\citenamefont {Tomioka}\ \emph {et~al.}(2007)\citenamefont
  {Tomioka}, \citenamefont {Mohan}, \citenamefont {Noborisaka}, \citenamefont
  {Hara}, \citenamefont {Motohisa},\ and\ \citenamefont {Fukui}}]{Tomioka07a}%
  \BibitemOpen
  \bibfield  {author} {\bibinfo {author} {\bibfnamefont {K.}~\bibnamefont
  {Tomioka}}, \bibinfo {author} {\bibfnamefont {P.}~\bibnamefont {Mohan}},
  \bibinfo {author} {\bibfnamefont {J.}~\bibnamefont {Noborisaka}}, \bibinfo
  {author} {\bibfnamefont {S.}~\bibnamefont {Hara}}, \bibinfo {author}
  {\bibfnamefont {J.}~\bibnamefont {Motohisa}}, \ and\ \bibinfo {author}
  {\bibfnamefont {T.}~\bibnamefont {Fukui}},\ }\href {\doibase
  http://dx.doi.org/10.1016/j.jcrysgro.2006.10.183} {\bibfield  {journal}
  {\bibinfo  {journal} {J. Cryst. Growth}\ }\textbf {\bibinfo {volume} {298}},\
  \bibinfo {pages} {644} (\bibinfo {year} {2007})}\BibitemShut {NoStop}%
\bibitem [{\citenamefont {Thelander}\ \emph {et~al.}(2008)\citenamefont
  {Thelander}, \citenamefont {Rehnstedt}, \citenamefont {Froberg},
  \citenamefont {Lind}, \citenamefont {Martensson}, \citenamefont {Caroff},
  \citenamefont {Lowgren}, \citenamefont {Ohlsson}, \citenamefont {Samuelson},\
  and\ \citenamefont {Wernersson}}]{Thelander08}%
  \BibitemOpen
  \bibfield  {author} {\bibinfo {author} {\bibfnamefont {C.}~\bibnamefont
  {Thelander}}, \bibinfo {author} {\bibfnamefont {C.}~\bibnamefont
  {Rehnstedt}}, \bibinfo {author} {\bibfnamefont {L.~E.}\ \bibnamefont
  {Froberg}}, \bibinfo {author} {\bibfnamefont {E.}~\bibnamefont {Lind}},
  \bibinfo {author} {\bibfnamefont {T.}~\bibnamefont {Martensson}}, \bibinfo
  {author} {\bibfnamefont {P.}~\bibnamefont {Caroff}}, \bibinfo {author}
  {\bibfnamefont {T.}~\bibnamefont {Lowgren}}, \bibinfo {author} {\bibfnamefont
  {B.~J.}\ \bibnamefont {Ohlsson}}, \bibinfo {author} {\bibfnamefont
  {L.}~\bibnamefont {Samuelson}}, \ and\ \bibinfo {author} {\bibfnamefont
  {L.-E.}\ \bibnamefont {Wernersson}},\ }\href {\doibase
  10.1109/TED.2008.2005151} {\bibfield  {journal} {\bibinfo  {journal}
  {Electron Devices, IEEE Transactions on}\ }\textbf {\bibinfo {volume} {55}},\
  \bibinfo {pages} {3030} (\bibinfo {year} {2008})}\BibitemShut {NoStop}%
\bibitem [{\citenamefont {Burke}\ \emph {et~al.}(2015)\citenamefont {Burke},
  \citenamefont {Carrad}, \citenamefont {Gluschke}, \citenamefont {Storm},
  \citenamefont {Fahlvik~Svensson}, \citenamefont {Linke}, \citenamefont
  {Samuelson},\ and\ \citenamefont {Micolich}}]{Burke15}%
  \BibitemOpen
  \bibfield  {author} {\bibinfo {author} {\bibfnamefont {A.~M.}\ \bibnamefont
  {Burke}}, \bibinfo {author} {\bibfnamefont {D.~J.}\ \bibnamefont {Carrad}},
  \bibinfo {author} {\bibfnamefont {J.~G.}\ \bibnamefont {Gluschke}}, \bibinfo
  {author} {\bibfnamefont {K.}~\bibnamefont {Storm}}, \bibinfo {author}
  {\bibfnamefont {S.}~\bibnamefont {Fahlvik~Svensson}}, \bibinfo {author}
  {\bibfnamefont {H.}~\bibnamefont {Linke}}, \bibinfo {author} {\bibfnamefont
  {L.}~\bibnamefont {Samuelson}}, \ and\ \bibinfo {author} {\bibfnamefont
  {A.~P.}\ \bibnamefont {Micolich}},\ }\href {\doibase 10.1021/nl5043243}
  {\bibfield  {journal} {\bibinfo  {journal} {Nano Letters}\ }\textbf {\bibinfo
  {volume} {15}},\ \bibinfo {pages} {2836} (\bibinfo {year}
  {2015})}\BibitemShut {NoStop}%
\bibitem [{\citenamefont {Fasth}\ \emph {et~al.}(2005)\citenamefont {Fasth},
  \citenamefont {Fuhrer}, \citenamefont {Bjork},\ and\ \citenamefont
  {Samuelson}}]{Fasth05a}%
  \BibitemOpen
  \bibfield  {author} {\bibinfo {author} {\bibfnamefont {C.}~\bibnamefont
  {Fasth}}, \bibinfo {author} {\bibfnamefont {A.}~\bibnamefont {Fuhrer}},
  \bibinfo {author} {\bibfnamefont {M.~T.}\ \bibnamefont {Bjork}}, \ and\
  \bibinfo {author} {\bibfnamefont {L.}~\bibnamefont {Samuelson}},\ }\href
  {\doibase http://dx.doi.org/10.1021/nl050850i} {\bibfield  {journal}
  {\bibinfo  {journal} {Nanoletters}\ }\textbf {\bibinfo {volume} {5}},\
  \bibinfo {pages} {1487} (\bibinfo {year} {2005})}\BibitemShut {NoStop}%
\bibitem [{\citenamefont {Fuhrer}\ \emph {et~al.}(2007)\citenamefont {Fuhrer},
  \citenamefont {Fr\"oberg}, \citenamefont {Pedersen}, \citenamefont {Larsson},
  \citenamefont {Wacker}, \citenamefont {Pistol},\ and\ \citenamefont
  {Samuelson}}]{Fuhrer07a}%
  \BibitemOpen
  \bibfield  {author} {\bibinfo {author} {\bibfnamefont {A.}~\bibnamefont
  {Fuhrer}}, \bibinfo {author} {\bibfnamefont {L.~E.}\ \bibnamefont
  {Fr\"oberg}}, \bibinfo {author} {\bibfnamefont {J.~N.}\ \bibnamefont
  {Pedersen}}, \bibinfo {author} {\bibfnamefont {M.~W.}\ \bibnamefont
  {Larsson}}, \bibinfo {author} {\bibfnamefont {A.}~\bibnamefont {Wacker}},
  \bibinfo {author} {\bibfnamefont {M.-E.}\ \bibnamefont {Pistol}}, \ and\
  \bibinfo {author} {\bibfnamefont {L.}~\bibnamefont {Samuelson}},\ }\href
  {\doibase 10.1021/nl061913f} {\bibfield  {journal} {\bibinfo  {journal} {Nano
  Letters}\ }\textbf {\bibinfo {volume} {7}},\ \bibinfo {pages} {243} (\bibinfo
  {year} {2007})}\BibitemShut {NoStop}%
\bibitem [{\citenamefont {Heedt}\ \emph {et~al.}(2016)\citenamefont {Heedt},
  \citenamefont {Manolescu}, \citenamefont {Nemnes}, \citenamefont {Prost},
  \citenamefont {Schubert}, \citenamefont {Gr\"utzmacher},\ and\ \citenamefont
  {Sch\"apers}}]{Heedt16b}%
  \BibitemOpen
  \bibfield  {author} {\bibinfo {author} {\bibfnamefont {S.}~\bibnamefont
  {Heedt}}, \bibinfo {author} {\bibfnamefont {A.}~\bibnamefont {Manolescu}},
  \bibinfo {author} {\bibfnamefont {G.~A.}\ \bibnamefont {Nemnes}}, \bibinfo
  {author} {\bibfnamefont {W.}~\bibnamefont {Prost}}, \bibinfo {author}
  {\bibfnamefont {J.}~\bibnamefont {Schubert}}, \bibinfo {author}
  {\bibfnamefont {D.}~\bibnamefont {Gr\"utzmacher}}, \ and\ \bibinfo {author}
  {\bibfnamefont {T.}~\bibnamefont {Sch\"apers}},\ }\href {\doibase
  10.1021/acs.nanolett.6b01840} {\bibfield  {journal} {\bibinfo  {journal}
  {Nano Letters}\ }\textbf {\bibinfo {volume} {16}},\ \bibinfo {pages} {4569}
  (\bibinfo {year} {2016})}\BibitemShut {NoStop}%
\bibitem [{\citenamefont {Heedt}\ \emph {et~al.}(2017)\citenamefont {Heedt},
  \citenamefont {Ziani}, \citenamefont {Cr{\'e}pin}, \citenamefont {Prost},
  \citenamefont {Schubert}, \citenamefont {Gr{\"u}tzmacher}, \citenamefont
  {Trauzettel}, \citenamefont {Sch{\"a}pers} \emph {et~al.}}]{Heedt17}%
  \BibitemOpen
  \bibfield  {author} {\bibinfo {author} {\bibfnamefont {S.}~\bibnamefont
  {Heedt}}, \bibinfo {author} {\bibfnamefont {N.~T.}\ \bibnamefont {Ziani}},
  \bibinfo {author} {\bibfnamefont {F.}~\bibnamefont {Cr{\'e}pin}}, \bibinfo
  {author} {\bibfnamefont {W.}~\bibnamefont {Prost}}, \bibinfo {author}
  {\bibfnamefont {J.}~\bibnamefont {Schubert}}, \bibinfo {author}
  {\bibfnamefont {D.}~\bibnamefont {Gr{\"u}tzmacher}}, \bibinfo {author}
  {\bibfnamefont {B.}~\bibnamefont {Trauzettel}}, \bibinfo {author}
  {\bibfnamefont {T.}~\bibnamefont {Sch{\"a}pers}},  \emph {et~al.},\ }\href
  {\doibase 10.1038/nphys4070} {\bibfield  {journal} {\bibinfo  {journal}
  {Nature Physics}\ }\textbf {\bibinfo {volume} {13}},\ \bibinfo {pages} {563}
  (\bibinfo {year} {2017})}\BibitemShut {NoStop}%
\bibitem [{\citenamefont {Iorio}\ \emph {et~al.}(2019)\citenamefont {Iorio},
  \citenamefont {Rocci}, \citenamefont {Bours}, \citenamefont {Carrega},
  \citenamefont {Zannier}, \citenamefont {Sorba}, \citenamefont {Roddaro},
  \citenamefont {Giazotto},\ and\ \citenamefont {Strambini}}]{Iorio18}%
  \BibitemOpen
  \bibfield  {author} {\bibinfo {author} {\bibfnamefont {A.}~\bibnamefont
  {Iorio}}, \bibinfo {author} {\bibfnamefont {M.}~\bibnamefont {Rocci}},
  \bibinfo {author} {\bibfnamefont {L.}~\bibnamefont {Bours}}, \bibinfo
  {author} {\bibfnamefont {M.}~\bibnamefont {Carrega}}, \bibinfo {author}
  {\bibfnamefont {V.}~\bibnamefont {Zannier}}, \bibinfo {author} {\bibfnamefont
  {L.}~\bibnamefont {Sorba}}, \bibinfo {author} {\bibfnamefont
  {S.}~\bibnamefont {Roddaro}}, \bibinfo {author} {\bibfnamefont
  {F.}~\bibnamefont {Giazotto}}, \ and\ \bibinfo {author} {\bibfnamefont
  {E.}~\bibnamefont {Strambini}},\ }\href {\doibase
  10.1021/acs.nanolett.8b02828} {\bibfield  {journal} {\bibinfo  {journal}
  {Nano Letters}\ }\textbf {\bibinfo {volume} {19}},\ \bibinfo {pages} {652}
  (\bibinfo {year} {2019})}\BibitemShut {NoStop}%
\bibitem [{\citenamefont {Bordoloi}\ \emph {et~al.}(2020)\citenamefont
  {Bordoloi}, \citenamefont {Zannier}, \citenamefont {Sorba}, \citenamefont
  {Sch{\"o}nenberger},\ and\ \citenamefont {Baumgartner}}]{Bordoloi20}%
  \BibitemOpen
  \bibfield  {author} {\bibinfo {author} {\bibfnamefont {A.}~\bibnamefont
  {Bordoloi}}, \bibinfo {author} {\bibfnamefont {V.}~\bibnamefont {Zannier}},
  \bibinfo {author} {\bibfnamefont {L.}~\bibnamefont {Sorba}}, \bibinfo
  {author} {\bibfnamefont {C.}~\bibnamefont {Sch{\"o}nenberger}}, \ and\
  \bibinfo {author} {\bibfnamefont {A.}~\bibnamefont {Baumgartner}},\ }\href
  {\doibase https://doi.org/10.1038/s42005-020-00405-2} {\bibfield  {journal}
  {\bibinfo  {journal} {Communications Physics}\ }\textbf {\bibinfo {volume}
  {3}},\ \bibinfo {pages} {1} (\bibinfo {year} {2020})}\BibitemShut {NoStop}%
\bibitem [{\citenamefont {Mourik}\ \emph {et~al.}(2012)\citenamefont {Mourik},
  \citenamefont {Zuo}, \citenamefont {Frolov}, \citenamefont {Plissard},
  \citenamefont {Bakkers},\ and\ \citenamefont {Kouwenhoven}}]{Mourik12}%
  \BibitemOpen
  \bibfield  {author} {\bibinfo {author} {\bibfnamefont {V.}~\bibnamefont
  {Mourik}}, \bibinfo {author} {\bibfnamefont {K.}~\bibnamefont {Zuo}},
  \bibinfo {author} {\bibfnamefont {S.~M.}\ \bibnamefont {Frolov}}, \bibinfo
  {author} {\bibfnamefont {S.~R.}\ \bibnamefont {Plissard}}, \bibinfo {author}
  {\bibfnamefont {E.~P. A.~M.}\ \bibnamefont {Bakkers}}, \ and\ \bibinfo
  {author} {\bibfnamefont {L.~P.}\ \bibnamefont {Kouwenhoven}},\ }\href
  {\doibase 10.1126/science.1222360} {\bibfield  {journal} {\bibinfo  {journal}
  {Science}\ }\textbf {\bibinfo {volume} {336}},\ \bibinfo {pages} {1003}
  (\bibinfo {year} {2012})}\BibitemShut {NoStop}%
\bibitem [{\citenamefont {Aguado}(2017)}]{Aguado17}%
  \BibitemOpen
  \bibfield  {author} {\bibinfo {author} {\bibfnamefont {R.}~\bibnamefont
  {Aguado}},\ }\href {\doibase 10.1393/ncr/i2017-10141-9} {\bibfield  {journal}
  {\bibinfo  {journal} {Rivista del Nuova Cimento}\ }\textbf {\bibinfo {volume}
  {40}},\ \bibinfo {pages} {523} (\bibinfo {year} {2017})}\BibitemShut
  {NoStop}%
\bibitem [{\citenamefont {Doh}\ \emph {et~al.}(2005)\citenamefont {Doh},
  \citenamefont {van Dam}, \citenamefont {Roest}, \citenamefont {Bakkers},
  \citenamefont {Kouwenhoven},\ and\ \citenamefont {Franceschi}}]{Doh05}%
  \BibitemOpen
  \bibfield  {author} {\bibinfo {author} {\bibfnamefont {Y.-J.}\ \bibnamefont
  {Doh}}, \bibinfo {author} {\bibfnamefont {J.~A.}\ \bibnamefont {van Dam}},
  \bibinfo {author} {\bibfnamefont {A.~L.}\ \bibnamefont {Roest}}, \bibinfo
  {author} {\bibfnamefont {E.~P. A.~M.}\ \bibnamefont {Bakkers}}, \bibinfo
  {author} {\bibfnamefont {L.~P.}\ \bibnamefont {Kouwenhoven}}, \ and\ \bibinfo
  {author} {\bibfnamefont {S.~D.}\ \bibnamefont {Franceschi}},\ }\href
  {\doibase DOI: 10.1126/science.1113523} {\bibfield  {journal} {\bibinfo
  {journal} {Science}\ }\textbf {\bibinfo {volume} {309}},\ \bibinfo {pages}
  {272} (\bibinfo {year} {2005})}\BibitemShut {NoStop}%
\bibitem [{\citenamefont {G\"unel}\ \emph {et~al.}(2012)\citenamefont
  {G\"unel}, \citenamefont {Batov}, \citenamefont {Hardtdegen}, \citenamefont
  {Sladek}, \citenamefont {Winden}, \citenamefont {Weis}, \citenamefont
  {Panaitov}, \citenamefont {Gr\"utzmacher},\ and\ \citenamefont
  {Sch\"apers}}]{Guenel12}%
  \BibitemOpen
  \bibfield  {author} {\bibinfo {author} {\bibfnamefont {H.~Y.}\ \bibnamefont
  {G\"unel}}, \bibinfo {author} {\bibfnamefont {I.~E.}\ \bibnamefont {Batov}},
  \bibinfo {author} {\bibfnamefont {H.}~\bibnamefont {Hardtdegen}}, \bibinfo
  {author} {\bibfnamefont {K.}~\bibnamefont {Sladek}}, \bibinfo {author}
  {\bibfnamefont {A.}~\bibnamefont {Winden}}, \bibinfo {author} {\bibfnamefont
  {K.}~\bibnamefont {Weis}}, \bibinfo {author} {\bibfnamefont {G.}~\bibnamefont
  {Panaitov}}, \bibinfo {author} {\bibfnamefont {D.}~\bibnamefont
  {Gr\"utzmacher}}, \ and\ \bibinfo {author} {\bibfnamefont {T.}~\bibnamefont
  {Sch\"apers}},\ }\href {\doibase http://dx.doi.org/10.1063/1.4745024}
  {\bibfield  {journal} {\bibinfo  {journal} {Journal of Applied Physics}\
  }\textbf {\bibinfo {volume} {112}},\ \bibinfo {pages} {034316} (\bibinfo
  {year} {2012})}\BibitemShut {NoStop}%
\bibitem [{\citenamefont {Larsen}\ \emph {et~al.}(2015)\citenamefont {Larsen},
  \citenamefont {Petersson}, \citenamefont {Kuemmeth}, \citenamefont
  {Jespersen}, \citenamefont {Krogstrup}, \citenamefont {Nyg\aa{}rd},\ and\
  \citenamefont {Marcus}}]{Larsen15}%
  \BibitemOpen
  \bibfield  {author} {\bibinfo {author} {\bibfnamefont {T.~W.}\ \bibnamefont
  {Larsen}}, \bibinfo {author} {\bibfnamefont {K.~D.}\ \bibnamefont
  {Petersson}}, \bibinfo {author} {\bibfnamefont {F.}~\bibnamefont {Kuemmeth}},
  \bibinfo {author} {\bibfnamefont {T.~S.}\ \bibnamefont {Jespersen}}, \bibinfo
  {author} {\bibfnamefont {P.}~\bibnamefont {Krogstrup}}, \bibinfo {author}
  {\bibfnamefont {J.}~\bibnamefont {Nyg\aa{}rd}}, \ and\ \bibinfo {author}
  {\bibfnamefont {C.~M.}\ \bibnamefont {Marcus}},\ }\href {\doibase
  10.1103/PhysRevLett.115.127001} {\bibfield  {journal} {\bibinfo  {journal}
  {Phys. Rev. Lett.}\ }\textbf {\bibinfo {volume} {115}},\ \bibinfo {pages}
  {127001} (\bibinfo {year} {2015})}\BibitemShut {NoStop}%
\bibitem [{\citenamefont {de~Lange}\ \emph {et~al.}(2015)\citenamefont
  {de~Lange}, \citenamefont {van Heck}, \citenamefont {Bruno}, \citenamefont
  {van Woerkom}, \citenamefont {Geresdi}, \citenamefont {Plissard},
  \citenamefont {Bakkers}, \citenamefont {Akhmerov},\ and\ \citenamefont
  {DiCarlo}}]{deLange15}%
  \BibitemOpen
  \bibfield  {author} {\bibinfo {author} {\bibfnamefont {G.}~\bibnamefont
  {de~Lange}}, \bibinfo {author} {\bibfnamefont {B.}~\bibnamefont {van Heck}},
  \bibinfo {author} {\bibfnamefont {A.}~\bibnamefont {Bruno}}, \bibinfo
  {author} {\bibfnamefont {D.~J.}\ \bibnamefont {van Woerkom}}, \bibinfo
  {author} {\bibfnamefont {A.}~\bibnamefont {Geresdi}}, \bibinfo {author}
  {\bibfnamefont {S.~R.}\ \bibnamefont {Plissard}}, \bibinfo {author}
  {\bibfnamefont {E.~P. A.~M.}\ \bibnamefont {Bakkers}}, \bibinfo {author}
  {\bibfnamefont {A.~R.}\ \bibnamefont {Akhmerov}}, \ and\ \bibinfo {author}
  {\bibfnamefont {L.}~\bibnamefont {DiCarlo}},\ }\href {\doibase
  10.1103/PhysRevLett.115.127002} {\bibfield  {journal} {\bibinfo  {journal}
  {Phys. Rev. Lett.}\ }\textbf {\bibinfo {volume} {115}},\ \bibinfo {pages}
  {127002} (\bibinfo {year} {2015})}\BibitemShut {NoStop}%
\bibitem [{\citenamefont {Zazunov}\ \emph {et~al.}(2003)\citenamefont
  {Zazunov}, \citenamefont {S.}, \citenamefont {Shumeiko}, \citenamefont
  {Bratus}, \citenamefont {Lantz},\ and\ \citenamefont {Wendin}}]{Zazunov03}%
  \BibitemOpen
  \bibfield  {author} {\bibinfo {author} {\bibfnamefont {A.}~\bibnamefont
  {Zazunov}}, \bibinfo {author} {\bibfnamefont {V.}~\bibnamefont {S.}},
  \bibinfo {author} {\bibnamefont {Shumeiko}}, \bibinfo {author} {\bibfnamefont
  {E.~N.}\ \bibnamefont {Bratus}}, \bibinfo {author} {\bibfnamefont
  {J.}~\bibnamefont {Lantz}}, \ and\ \bibinfo {author} {\bibfnamefont
  {G.}~\bibnamefont {Wendin}},\ }\href {\doibase 10.1103/PhysRevLett.90.087003}
  {\bibfield  {journal} {\bibinfo  {journal} {Phys. Rev. Lett.}\ }\textbf
  {\bibinfo {volume} {90}},\ \bibinfo {pages} {087003} (\bibinfo {year}
  {2003})}\BibitemShut {NoStop}%
\bibitem [{\citenamefont {Wieder}(1974)}]{Wieder74}%
  \BibitemOpen
  \bibfield  {author} {\bibinfo {author} {\bibfnamefont {H.~H.}\ \bibnamefont
  {Wieder}},\ }\href {\doibase 10.1063/1.1655441} {\bibfield  {journal}
  {\bibinfo  {journal} {Applied Physics Letters}\ }\textbf {\bibinfo {volume}
  {25}},\ \bibinfo {pages} {206} (\bibinfo {year} {1974})}\BibitemShut
  {NoStop}%
\bibitem [{\citenamefont {Sladek}\ \emph {et~al.}(2010)\citenamefont {Sladek},
  \citenamefont {Klinger}, \citenamefont {Wensorra}, \citenamefont {Akabori},
  \citenamefont {Hardtdegen},\ and\ \citenamefont {Gr\"utzmacher}}]{Sladek10a}%
  \BibitemOpen
  \bibfield  {author} {\bibinfo {author} {\bibfnamefont {K.}~\bibnamefont
  {Sladek}}, \bibinfo {author} {\bibfnamefont {V.}~\bibnamefont {Klinger}},
  \bibinfo {author} {\bibfnamefont {J.}~\bibnamefont {Wensorra}}, \bibinfo
  {author} {\bibfnamefont {M.}~\bibnamefont {Akabori}}, \bibinfo {author}
  {\bibfnamefont {H.}~\bibnamefont {Hardtdegen}}, \ and\ \bibinfo {author}
  {\bibfnamefont {D.}~\bibnamefont {Gr\"utzmacher}},\ }\href {\doibase
  10.1016/j.jcrysgro.2009.11.026} {\bibfield  {journal} {\bibinfo  {journal}
  {Journal of Crystal Growth}\ }\textbf {\bibinfo {volume} {312}},\ \bibinfo
  {pages} {635 } (\bibinfo {year} {2010})}\BibitemShut {NoStop}%
\bibitem [{\citenamefont {Wirths}\ \emph {et~al.}(2011)\citenamefont {Wirths},
  \citenamefont {Weis}, \citenamefont {Winden}, \citenamefont {Sladek},
  \citenamefont {Volk}, \citenamefont {Alagha}, \citenamefont {Weirich},
  \citenamefont {von~der Ahe}, \citenamefont {Hardtdegen}, \citenamefont
  {L\"{u}th}, \citenamefont {Demarina}, \citenamefont {Gr\"{u}tzmacher},\ and\
  \citenamefont {Sch\"{a}pers}}]{Wirths11}%
  \BibitemOpen
  \bibfield  {author} {\bibinfo {author} {\bibfnamefont {S.}~\bibnamefont
  {Wirths}}, \bibinfo {author} {\bibfnamefont {K.}~\bibnamefont {Weis}},
  \bibinfo {author} {\bibfnamefont {A.}~\bibnamefont {Winden}}, \bibinfo
  {author} {\bibfnamefont {K.}~\bibnamefont {Sladek}}, \bibinfo {author}
  {\bibfnamefont {C.}~\bibnamefont {Volk}}, \bibinfo {author} {\bibfnamefont
  {S.}~\bibnamefont {Alagha}}, \bibinfo {author} {\bibfnamefont {T.~E.}\
  \bibnamefont {Weirich}}, \bibinfo {author} {\bibfnamefont {M.}~\bibnamefont
  {von~der Ahe}}, \bibinfo {author} {\bibfnamefont {H.}~\bibnamefont
  {Hardtdegen}}, \bibinfo {author} {\bibfnamefont {H.}~\bibnamefont
  {L\"{u}th}}, \bibinfo {author} {\bibfnamefont {N.}~\bibnamefont {Demarina}},
  \bibinfo {author} {\bibfnamefont {D.}~\bibnamefont {Gr\"{u}tzmacher}}, \ and\
  \bibinfo {author} {\bibfnamefont {T.}~\bibnamefont {Sch\"{a}pers}},\ }\href
  {\doibase 10.1063/1.3631026} {\bibfield  {journal} {\bibinfo  {journal}
  {Journal of Applied Physics}\ }\textbf {\bibinfo {volume} {110}},\ \bibinfo
  {eid} {053709} (\bibinfo {year} {2011})}\BibitemShut {NoStop}%
\bibitem [{\citenamefont {Ghoneim}\ \emph {et~al.}(2012)\citenamefont
  {Ghoneim}, \citenamefont {Mensch}, \citenamefont {Schmid}, \citenamefont
  {Bessire}, \citenamefont {Rhyner}, \citenamefont {Schenk}, \citenamefont
  {Rettner}, \citenamefont {Karg}, \citenamefont {Moselund}, \citenamefont
  {Riel},\ and\ \citenamefont {Björk}}]{Ghoneim12}%
  \BibitemOpen
  \bibfield  {author} {\bibinfo {author} {\bibfnamefont {H.}~\bibnamefont
  {Ghoneim}}, \bibinfo {author} {\bibfnamefont {P.}~\bibnamefont {Mensch}},
  \bibinfo {author} {\bibfnamefont {H.}~\bibnamefont {Schmid}}, \bibinfo
  {author} {\bibfnamefont {C.~D.}\ \bibnamefont {Bessire}}, \bibinfo {author}
  {\bibfnamefont {R.}~\bibnamefont {Rhyner}}, \bibinfo {author} {\bibfnamefont
  {A.}~\bibnamefont {Schenk}}, \bibinfo {author} {\bibfnamefont
  {C.}~\bibnamefont {Rettner}}, \bibinfo {author} {\bibfnamefont
  {S.}~\bibnamefont {Karg}}, \bibinfo {author} {\bibfnamefont {K.~E.}\
  \bibnamefont {Moselund}}, \bibinfo {author} {\bibfnamefont {H.}~\bibnamefont
  {Riel}}, \ and\ \bibinfo {author} {\bibfnamefont {M.~T.}\ \bibnamefont
  {Björk}},\ }\href {\doibase 10.1088/0957-4484/23/50/505708} {\bibfield
  {journal} {\bibinfo  {journal} {Nanotechnology}\ }\textbf {\bibinfo {volume}
  {23}},\ \bibinfo {pages} {505708} (\bibinfo {year} {2012})}\BibitemShut
  {NoStop}%
\bibitem [{\citenamefont {Dimakis}\ \emph {et~al.}(2013)\citenamefont
  {Dimakis}, \citenamefont {Ramsteiner}, \citenamefont {Huang}, \citenamefont
  {Trampert}, \citenamefont {Davydok}, \citenamefont {Biermanns}, \citenamefont
  {Pietsch}, \citenamefont {Riechert},\ and\ \citenamefont
  {Geelhaar}}]{Dimakis13}%
  \BibitemOpen
  \bibfield  {author} {\bibinfo {author} {\bibfnamefont {E.}~\bibnamefont
  {Dimakis}}, \bibinfo {author} {\bibfnamefont {M.}~\bibnamefont {Ramsteiner}},
  \bibinfo {author} {\bibfnamefont {C.-N.}\ \bibnamefont {Huang}}, \bibinfo
  {author} {\bibfnamefont {A.}~\bibnamefont {Trampert}}, \bibinfo {author}
  {\bibfnamefont {A.}~\bibnamefont {Davydok}}, \bibinfo {author} {\bibfnamefont
  {A.}~\bibnamefont {Biermanns}}, \bibinfo {author} {\bibfnamefont
  {U.}~\bibnamefont {Pietsch}}, \bibinfo {author} {\bibfnamefont
  {H.}~\bibnamefont {Riechert}}, \ and\ \bibinfo {author} {\bibfnamefont
  {L.}~\bibnamefont {Geelhaar}},\ }\href {\doibase 10.1063/1.4824344}
  {\bibfield  {journal} {\bibinfo  {journal} {Applied Physics Letters}\
  }\textbf {\bibinfo {volume} {103}},\ \bibinfo {pages} {143121} (\bibinfo
  {year} {2013})}\BibitemShut {NoStop}%
\bibitem [{\citenamefont {Park}\ \emph {et~al.}(2015)\citenamefont {Park},
  \citenamefont {Jeon}, \citenamefont {Lee}, \citenamefont {Lee}, \citenamefont
  {Song}, \citenamefont {Kim}, \citenamefont {Noh}, \citenamefont {Leem},\ and\
  \citenamefont {Kim}}]{Park15}%
  \BibitemOpen
  \bibfield  {author} {\bibinfo {author} {\bibfnamefont {D.~W.}\ \bibnamefont
  {Park}}, \bibinfo {author} {\bibfnamefont {S.~G.}\ \bibnamefont {Jeon}},
  \bibinfo {author} {\bibfnamefont {C.-R.}\ \bibnamefont {Lee}}, \bibinfo
  {author} {\bibfnamefont {S.~J.}\ \bibnamefont {Lee}}, \bibinfo {author}
  {\bibfnamefont {J.~Y.}\ \bibnamefont {Song}}, \bibinfo {author}
  {\bibfnamefont {J.~O.}\ \bibnamefont {Kim}}, \bibinfo {author} {\bibfnamefont
  {S.~K.}\ \bibnamefont {Noh}}, \bibinfo {author} {\bibfnamefont {J.-Y.}\
  \bibnamefont {Leem}}, \ and\ \bibinfo {author} {\bibfnamefont {J.~S.}\
  \bibnamefont {Kim}},\ }\href {\doibase 10.1038/srep16652} {\bibfield
  {journal} {\bibinfo  {journal} {Scientific Reports}\ }\textbf {\bibinfo
  {volume} {5}} (\bibinfo {year} {2015}),\ 10.1038/srep16652}\BibitemShut
  {NoStop}%
\bibitem [{\citenamefont {Sankaran}(1980)}]{Sankaran80}%
  \BibitemOpen
  \bibfield  {author} {\bibinfo {author} {\bibfnamefont {R.}~\bibnamefont
  {Sankaran}},\ }\href@noop {} {\bibfield  {journal} {\bibinfo  {journal}
  {Journal of Crystal Growth}\ }\textbf {\bibinfo {volume} {50}},\ \bibinfo
  {pages} {859} (\bibinfo {year} {1980})}\BibitemShut {NoStop}%
\bibitem [{\citenamefont {Kamp}\ \emph {et~al.}(1994)\citenamefont {Kamp},
  \citenamefont {M\"orsch}, \citenamefont {Gr\"aber},\ and\ \citenamefont
  {L\"uth}}]{Kamp94}%
  \BibitemOpen
  \bibfield  {author} {\bibinfo {author} {\bibfnamefont {M.}~\bibnamefont
  {Kamp}}, \bibinfo {author} {\bibfnamefont {G.}~\bibnamefont {M\"orsch}},
  \bibinfo {author} {\bibfnamefont {J.}~\bibnamefont {Gr\"aber}}, \ and\
  \bibinfo {author} {\bibfnamefont {H.}~\bibnamefont {L\"uth}},\ }\href
  {\doibase 10.1063/1.357660} {\bibfield  {journal} {\bibinfo  {journal}
  {Journal of Applied Physics}\ }\textbf {\bibinfo {volume} {76}},\ \bibinfo
  {pages} {1974} (\bibinfo {year} {1994})}\BibitemShut {NoStop}%
\bibitem [{\citenamefont {Bennett}\ \emph {et~al.}(2003)\citenamefont
  {Bennett}, \citenamefont {Magno},\ and\ \citenamefont
  {Papanicolaou}}]{Bennett03}%
  \BibitemOpen
  \bibfield  {author} {\bibinfo {author} {\bibfnamefont {B.~R.}\ \bibnamefont
  {Bennett}}, \bibinfo {author} {\bibfnamefont {R.}~\bibnamefont {Magno}}, \
  and\ \bibinfo {author} {\bibfnamefont {N.}~\bibnamefont {Papanicolaou}},\
  }\href {\doibase https://doi.org/10.1016/S0022-0248(02)02186-3} {\bibfield
  {journal} {\bibinfo  {journal} {Journal of crystal growth}\ }\textbf
  {\bibinfo {volume} {251}},\ \bibinfo {pages} {532} (\bibinfo {year}
  {2003})}\BibitemShut {NoStop}%
\bibitem [{\citenamefont {Wallentin}\ and\ \citenamefont
  {Borgstr{\"o}m}(2011)}]{Wallentin11a}%
  \BibitemOpen
  \bibfield  {author} {\bibinfo {author} {\bibfnamefont {J.}~\bibnamefont
  {Wallentin}}\ and\ \bibinfo {author} {\bibfnamefont {M.~T.}\ \bibnamefont
  {Borgstr{\"o}m}},\ }\href {\doibase https://doi.org/10.1557/jmr.2011.214}
  {\bibfield  {journal} {\bibinfo  {journal} {Journal of Materials Research}\
  }\textbf {\bibinfo {volume} {26}},\ \bibinfo {pages} {2142} (\bibinfo {year}
  {2011})}\BibitemShut {NoStop}%
\bibitem [{\citenamefont {Orr{\`u}}\ \emph {et~al.}(2016)\citenamefont
  {Orr{\`u}}, \citenamefont {Repiso}, \citenamefont {Carapezzi}, \citenamefont
  {Henning}, \citenamefont {Roddaro}, \citenamefont {Franciosi}, \citenamefont
  {Rosenwaks}, \citenamefont {Cavallini}, \citenamefont {Martelli},\ and\
  \citenamefont {Rubini}}]{Orru16}%
  \BibitemOpen
  \bibfield  {author} {\bibinfo {author} {\bibfnamefont {M.}~\bibnamefont
  {Orr{\`u}}}, \bibinfo {author} {\bibfnamefont {E.}~\bibnamefont {Repiso}},
  \bibinfo {author} {\bibfnamefont {S.}~\bibnamefont {Carapezzi}}, \bibinfo
  {author} {\bibfnamefont {A.}~\bibnamefont {Henning}}, \bibinfo {author}
  {\bibfnamefont {S.}~\bibnamefont {Roddaro}}, \bibinfo {author} {\bibfnamefont
  {A.}~\bibnamefont {Franciosi}}, \bibinfo {author} {\bibfnamefont
  {Y.}~\bibnamefont {Rosenwaks}}, \bibinfo {author} {\bibfnamefont
  {A.}~\bibnamefont {Cavallini}}, \bibinfo {author} {\bibfnamefont
  {F.}~\bibnamefont {Martelli}}, \ and\ \bibinfo {author} {\bibfnamefont
  {S.}~\bibnamefont {Rubini}},\ }\href {\doibase
  https://doi.org/10.1002/adfm.201504853} {\bibfield  {journal} {\bibinfo
  {journal} {Advanced Functional Materials}\ }\textbf {\bibinfo {volume}
  {26}},\ \bibinfo {pages} {2836} (\bibinfo {year} {2016})}\BibitemShut
  {NoStop}%
\bibitem [{\citenamefont {Goktas}\ \emph {et~al.}(2018)\citenamefont {Goktas},
  \citenamefont {Fiordaliso},\ and\ \citenamefont {LaPierre}}]{Goktas18}%
  \BibitemOpen
  \bibfield  {author} {\bibinfo {author} {\bibfnamefont {N.~I.}\ \bibnamefont
  {Goktas}}, \bibinfo {author} {\bibfnamefont {E.~M.}\ \bibnamefont
  {Fiordaliso}}, \ and\ \bibinfo {author} {\bibfnamefont {R.~R.}\ \bibnamefont
  {LaPierre}},\ }\href {\doibase 10.1088/1361-6528/aab6f1} {\bibfield
  {journal} {\bibinfo  {journal} {Nanotechnology}\ }\textbf {\bibinfo {volume}
  {29}},\ \bibinfo {pages} {234001} (\bibinfo {year} {2018})}\BibitemShut
  {NoStop}%
\bibitem [{\citenamefont {Hakkarainen}\ \emph {et~al.}(2019)\citenamefont
  {Hakkarainen}, \citenamefont {Rizzo~Piton}, \citenamefont {Fiordaliso},
  \citenamefont {Leshchenko}, \citenamefont {Koelling}, \citenamefont
  {Bettini}, \citenamefont {Vinicius Avan\ifmmode \mbox{\c{c}}\else~\c{c}\fi{}o
  Galeti}, \citenamefont {Koivusalo}, \citenamefont {Gobato}, \citenamefont
  {de~Giovanni~Rodrigues}, \citenamefont {Lupo}, \citenamefont {Koenraad},
  \citenamefont {Leite}, \citenamefont {Dubrovskii},\ and\ \citenamefont
  {Guina}}]{Hakkarainen19}%
  \BibitemOpen
  \bibfield  {author} {\bibinfo {author} {\bibfnamefont {T.}~\bibnamefont
  {Hakkarainen}}, \bibinfo {author} {\bibfnamefont {M.}~\bibnamefont
  {Rizzo~Piton}}, \bibinfo {author} {\bibfnamefont {E.~M.}\ \bibnamefont
  {Fiordaliso}}, \bibinfo {author} {\bibfnamefont {E.~D.}\ \bibnamefont
  {Leshchenko}}, \bibinfo {author} {\bibfnamefont {S.}~\bibnamefont
  {Koelling}}, \bibinfo {author} {\bibfnamefont {J.}~\bibnamefont {Bettini}},
  \bibinfo {author} {\bibfnamefont {H.}~\bibnamefont {Vinicius Avan\ifmmode
  \mbox{\c{c}}\else~\c{c}\fi{}o Galeti}}, \bibinfo {author} {\bibfnamefont
  {E.}~\bibnamefont {Koivusalo}}, \bibinfo {author} {\bibfnamefont {Y.~G.~a.}\
  \bibnamefont {Gobato}}, \bibinfo {author} {\bibfnamefont {A.}~\bibnamefont
  {de~Giovanni~Rodrigues}}, \bibinfo {author} {\bibfnamefont {D.}~\bibnamefont
  {Lupo}}, \bibinfo {author} {\bibfnamefont {P.~M.}\ \bibnamefont {Koenraad}},
  \bibinfo {author} {\bibfnamefont {E.~R.}\ \bibnamefont {Leite}}, \bibinfo
  {author} {\bibfnamefont {V.~G.}\ \bibnamefont {Dubrovskii}}, \ and\ \bibinfo
  {author} {\bibfnamefont {M.}~\bibnamefont {Guina}},\ }\href {\doibase
  10.1103/PhysRevMaterials.3.086001} {\bibfield  {journal} {\bibinfo  {journal}
  {Phys. Rev. Materials}\ }\textbf {\bibinfo {volume} {3}},\ \bibinfo {pages}
  {086001} (\bibinfo {year} {2019})}\BibitemShut {NoStop}%
\bibitem [{\citenamefont {G{\"u}sken}\ \emph {et~al.}(2019)\citenamefont
  {G{\"u}sken}, \citenamefont {Rieger}, \citenamefont {Mussler}, \citenamefont
  {Lepsa},\ and\ \citenamefont {Gr{\"u}tzmacher}}]{Guesken19}%
  \BibitemOpen
  \bibfield  {author} {\bibinfo {author} {\bibfnamefont {N.~A.}\ \bibnamefont
  {G{\"u}sken}}, \bibinfo {author} {\bibfnamefont {T.}~\bibnamefont {Rieger}},
  \bibinfo {author} {\bibfnamefont {G.}~\bibnamefont {Mussler}}, \bibinfo
  {author} {\bibfnamefont {M.~I.}\ \bibnamefont {Lepsa}}, \ and\ \bibinfo
  {author} {\bibfnamefont {D.}~\bibnamefont {Gr{\"u}tzmacher}},\ }\href
  {\doibase 10.1186/s11671-019-3004-0} {\bibfield  {journal} {\bibinfo
  {journal} {Nanoscale Research Letters}\ }\textbf {\bibinfo {volume} {14}},\
  \bibinfo {pages} {179} (\bibinfo {year} {2019})}\BibitemShut {NoStop}%
\bibitem [{\citenamefont {Koblm\"uller}\ \emph {et~al.}(2010)\citenamefont
  {Koblm\"uller}, \citenamefont {Hertenberger}, \citenamefont {Vizbaras},
  \citenamefont {Bichler}, \citenamefont {Bao}, \citenamefont {Zhang},\ and\
  \citenamefont {Abstreiter}}]{Koblmueller10}%
  \BibitemOpen
  \bibfield  {author} {\bibinfo {author} {\bibfnamefont {G.}~\bibnamefont
  {Koblm\"uller}}, \bibinfo {author} {\bibfnamefont {S.}~\bibnamefont
  {Hertenberger}}, \bibinfo {author} {\bibfnamefont {K.}~\bibnamefont
  {Vizbaras}}, \bibinfo {author} {\bibfnamefont {M.}~\bibnamefont {Bichler}},
  \bibinfo {author} {\bibfnamefont {F.}~\bibnamefont {Bao}}, \bibinfo {author}
  {\bibfnamefont {J.-P.}\ \bibnamefont {Zhang}}, \ and\ \bibinfo {author}
  {\bibfnamefont {G.}~\bibnamefont {Abstreiter}},\ }\href
  {http://stacks.iop.org/0957-4484/21/i=36/a=365602} {\bibfield  {journal}
  {\bibinfo  {journal} {Nanotechnology}\ }\textbf {\bibinfo {volume} {21}},\
  \bibinfo {pages} {365602} (\bibinfo {year} {2010})}\BibitemShut {NoStop}%
\bibitem [{\citenamefont {Perla}\ \emph {et~al.}(2021)\citenamefont {Perla},
  \citenamefont {Fonseka}, \citenamefont {Zellekens}, \citenamefont {Deacon},
  \citenamefont {Han}, \citenamefont {K{\"o}lzer}, \citenamefont
  {M{\"o}rstedt}, \citenamefont {Bennemann}, \citenamefont {Espiari},
  \citenamefont {Ishibashi} \emph {et~al.}}]{Perla21}%
  \BibitemOpen
  \bibfield  {author} {\bibinfo {author} {\bibfnamefont {P.}~\bibnamefont
  {Perla}}, \bibinfo {author} {\bibfnamefont {H.~A.}\ \bibnamefont {Fonseka}},
  \bibinfo {author} {\bibfnamefont {P.}~\bibnamefont {Zellekens}}, \bibinfo
  {author} {\bibfnamefont {R.}~\bibnamefont {Deacon}}, \bibinfo {author}
  {\bibfnamefont {Y.}~\bibnamefont {Han}}, \bibinfo {author} {\bibfnamefont
  {J.}~\bibnamefont {K{\"o}lzer}}, \bibinfo {author} {\bibfnamefont
  {T.}~\bibnamefont {M{\"o}rstedt}}, \bibinfo {author} {\bibfnamefont
  {B.}~\bibnamefont {Bennemann}}, \bibinfo {author} {\bibfnamefont
  {A.}~\bibnamefont {Espiari}}, \bibinfo {author} {\bibfnamefont
  {K.}~\bibnamefont {Ishibashi}},  \emph {et~al.},\ }\href {\doibase
  10.1039/d0na00999g} {\bibfield  {journal} {\bibinfo  {journal} {Nanoscale
  Advances}\ }\textbf {\bibinfo {volume} {3}},\ \bibinfo {pages} {1413}
  (\bibinfo {year} {2021})}\BibitemShut {NoStop}%
\bibitem [{\citenamefont {Rieger}\ \emph {et~al.}(2013)\citenamefont {Rieger},
  \citenamefont {Gr{\"u}tzmacher},\ and\ \citenamefont {Lepsa}}]{Rieger13a}%
  \BibitemOpen
  \bibfield  {author} {\bibinfo {author} {\bibfnamefont {T.}~\bibnamefont
  {Rieger}}, \bibinfo {author} {\bibfnamefont {D.}~\bibnamefont
  {Gr{\"u}tzmacher}}, \ and\ \bibinfo {author} {\bibfnamefont {M.~I.}\
  \bibnamefont {Lepsa}},\ }\href {\doibase
  https://doi.org/10.1002/pssr.201307229} {\bibfield  {journal} {\bibinfo
  {journal} {physica status solidi (RRL)--Rapid Research Letters}\ }\textbf
  {\bibinfo {volume} {7}},\ \bibinfo {pages} {840} (\bibinfo {year}
  {2013})}\BibitemShut {NoStop}%
\bibitem [{\citenamefont {Wixom}\ \emph {et~al.}(2004)\citenamefont {Wixom},
  \citenamefont {Rieth},\ and\ \citenamefont {Stringfellow}}]{Wixom04}%
  \BibitemOpen
  \bibfield  {author} {\bibinfo {author} {\bibfnamefont {R.}~\bibnamefont
  {Wixom}}, \bibinfo {author} {\bibfnamefont {L.}~\bibnamefont {Rieth}}, \ and\
  \bibinfo {author} {\bibfnamefont {G.}~\bibnamefont {Stringfellow}},\ }\href
  {\doibase https://doi.org/10.1016/j.jcrysgro.2004.05.077} {\bibfield
  {journal} {\bibinfo  {journal} {Journal of crystal growth}\ }\textbf
  {\bibinfo {volume} {269}},\ \bibinfo {pages} {276} (\bibinfo {year}
  {2004})}\BibitemShut {NoStop}%
\bibitem [{\citenamefont {S{\'a}far}\ \emph {et~al.}(1997)\citenamefont
  {S{\'a}far}, \citenamefont {Rodrigues}, \citenamefont {Cury}, \citenamefont
  {Chacham}, \citenamefont {Moreira}, \citenamefont {Freire},\ and\
  \citenamefont {de~Oliveira}}]{Safar97}%
  \BibitemOpen
  \bibfield  {author} {\bibinfo {author} {\bibfnamefont {G.~A.~M.}\
  \bibnamefont {S{\'a}far}}, \bibinfo {author} {\bibfnamefont {W.~N.}\
  \bibnamefont {Rodrigues}}, \bibinfo {author} {\bibfnamefont {L.~A.}\
  \bibnamefont {Cury}}, \bibinfo {author} {\bibfnamefont {H.}~\bibnamefont
  {Chacham}}, \bibinfo {author} {\bibfnamefont {M.~V.~B.}\ \bibnamefont
  {Moreira}}, \bibinfo {author} {\bibfnamefont {S.~L.~S.}\ \bibnamefont
  {Freire}}, \ and\ \bibinfo {author} {\bibfnamefont {A.~G.}\ \bibnamefont
  {de~Oliveira}},\ }\href {\doibase 10.1063/1.119597} {\bibfield  {journal}
  {\bibinfo  {journal} {Applied Physics Letters}\ }\textbf {\bibinfo {volume}
  {71}},\ \bibinfo {pages} {521} (\bibinfo {year} {1997})}\BibitemShut
  {NoStop}%
\bibitem [{\citenamefont {Neves}\ \emph {et~al.}(1998)\citenamefont {Neves},
  \citenamefont {Andrade}, \citenamefont {Rodrigues}, \citenamefont
  {S{\'a}far}, \citenamefont {Moreira},\ and\ \citenamefont
  {de~Oliveira}}]{Neves98}%
  \BibitemOpen
  \bibfield  {author} {\bibinfo {author} {\bibfnamefont {B.~R.~A.}\
  \bibnamefont {Neves}}, \bibinfo {author} {\bibfnamefont {M.~S.}\ \bibnamefont
  {Andrade}}, \bibinfo {author} {\bibfnamefont {W.~N.}\ \bibnamefont
  {Rodrigues}}, \bibinfo {author} {\bibfnamefont {G.~A.~M.}\ \bibnamefont
  {S{\'a}far}}, \bibinfo {author} {\bibfnamefont {M.~V.~B.}\ \bibnamefont
  {Moreira}}, \ and\ \bibinfo {author} {\bibfnamefont {A.~G.}\ \bibnamefont
  {de~Oliveira}},\ }\href {\doibase 10.1063/1.121160} {\bibfield  {journal}
  {\bibinfo  {journal} {Applied Physics Letters}\ }\textbf {\bibinfo {volume}
  {72}},\ \bibinfo {pages} {1712} (\bibinfo {year} {1998})}\BibitemShut
  {NoStop}%
\bibitem [{\citenamefont {Anyebe}\ \emph {et~al.}(2015)\citenamefont {Anyebe},
  \citenamefont {Rajpalke}, \citenamefont {Veal}, \citenamefont {Jin},
  \citenamefont {Wang},\ and\ \citenamefont {Zhuang}}]{Anyebe15}%
  \BibitemOpen
  \bibfield  {author} {\bibinfo {author} {\bibfnamefont {E.}~\bibnamefont
  {Anyebe}}, \bibinfo {author} {\bibfnamefont {M.~K.}\ \bibnamefont
  {Rajpalke}}, \bibinfo {author} {\bibfnamefont {T.~D.}\ \bibnamefont {Veal}},
  \bibinfo {author} {\bibfnamefont {C.}~\bibnamefont {Jin}}, \bibinfo {author}
  {\bibfnamefont {Z.}~\bibnamefont {Wang}}, \ and\ \bibinfo {author}
  {\bibfnamefont {Q.}~\bibnamefont {Zhuang}},\ }\href {\doibase
  https://doi.org/10.1007/s12274-014-0621-x} {\bibfield  {journal} {\bibinfo
  {journal} {Nano Research}\ }\textbf {\bibinfo {volume} {8}},\ \bibinfo
  {pages} {1309} (\bibinfo {year} {2015})}\BibitemShut {NoStop}%
\bibitem [{\citenamefont {Potts}\ \emph {et~al.}(2016)\citenamefont {Potts},
  \citenamefont {Friedl}, \citenamefont {Amaduzzi}, \citenamefont {Tang},
  \citenamefont {T{\"u}t{\"u}nc{\"u}oglu}, \citenamefont {Matteini},
  \citenamefont {Alarcon~Llad{\'o}}, \citenamefont {McIntyre},\ and\
  \citenamefont {Fontcuberta~i Morral}}]{Potts16}%
  \BibitemOpen
  \bibfield  {author} {\bibinfo {author} {\bibfnamefont {H.}~\bibnamefont
  {Potts}}, \bibinfo {author} {\bibfnamefont {M.}~\bibnamefont {Friedl}},
  \bibinfo {author} {\bibfnamefont {F.}~\bibnamefont {Amaduzzi}}, \bibinfo
  {author} {\bibfnamefont {K.}~\bibnamefont {Tang}}, \bibinfo {author}
  {\bibfnamefont {G.}~\bibnamefont {T{\"u}t{\"u}nc{\"u}oglu}}, \bibinfo
  {author} {\bibfnamefont {F.}~\bibnamefont {Matteini}}, \bibinfo {author}
  {\bibfnamefont {E.}~\bibnamefont {Alarcon~Llad{\'o}}}, \bibinfo {author}
  {\bibfnamefont {P.~C.}\ \bibnamefont {McIntyre}}, \ and\ \bibinfo {author}
  {\bibfnamefont {A.}~\bibnamefont {Fontcuberta~i Morral}},\ }\href {\doibase
  https://doi.org/10.1021/acs.nanolett.5b04367} {\bibfield  {journal} {\bibinfo
   {journal} {Nano letters}\ }\textbf {\bibinfo {volume} {16}},\ \bibinfo
  {pages} {637} (\bibinfo {year} {2016})}\BibitemShut {NoStop}%
\bibitem [{\citenamefont {Koelling}\ \emph {et~al.}(2017)\citenamefont
  {Koelling}, \citenamefont {Li}, \citenamefont {Cavalli}, \citenamefont
  {Assali}, \citenamefont {Car}, \citenamefont {Gazibegovic}, \citenamefont
  {Bakkers},\ and\ \citenamefont {Koenraad}}]{Koelling17}%
  \BibitemOpen
  \bibfield  {author} {\bibinfo {author} {\bibfnamefont {S.}~\bibnamefont
  {Koelling}}, \bibinfo {author} {\bibfnamefont {A.}~\bibnamefont {Li}},
  \bibinfo {author} {\bibfnamefont {A.}~\bibnamefont {Cavalli}}, \bibinfo
  {author} {\bibfnamefont {S.}~\bibnamefont {Assali}}, \bibinfo {author}
  {\bibfnamefont {D.}~\bibnamefont {Car}}, \bibinfo {author} {\bibfnamefont
  {S.}~\bibnamefont {Gazibegovic}}, \bibinfo {author} {\bibfnamefont {E.~P.
  A.~M.}\ \bibnamefont {Bakkers}}, \ and\ \bibinfo {author} {\bibfnamefont
  {P.~M.}\ \bibnamefont {Koenraad}},\ }\href {\doibase
  10.1021/acs.nanolett.6b03109} {\bibfield  {journal} {\bibinfo  {journal}
  {Nano Letters}\ }\textbf {\bibinfo {volume} {17}},\ \bibinfo {pages} {599}
  (\bibinfo {year} {2017})}\BibitemShut {NoStop}%
\bibitem [{\citenamefont {Hellman}\ \emph {et~al.}(2003)\citenamefont
  {Hellman}, \citenamefont {du~Rivage},\ and\ \citenamefont
  {Seidman}}]{Hellman03}%
  \BibitemOpen
  \bibfield  {author} {\bibinfo {author} {\bibfnamefont {O.~C.}\ \bibnamefont
  {Hellman}}, \bibinfo {author} {\bibfnamefont {J.~B.}\ \bibnamefont
  {du~Rivage}}, \ and\ \bibinfo {author} {\bibfnamefont {D.~N.}\ \bibnamefont
  {Seidman}},\ }\href {\doibase https://doi.org/10.1016/S0304-3991(02)00317-0}
  {\bibfield  {journal} {\bibinfo  {journal} {Ultramicroscopy}\ }\textbf
  {\bibinfo {volume} {95}},\ \bibinfo {pages} {199} (\bibinfo {year} {2003})},\
  \bibinfo {note} {iFES 2001}\BibitemShut {NoStop}%
\bibitem [{\citenamefont {Joyez}\ \emph {et~al.}(1999)\citenamefont {Joyez},
  \citenamefont {Vion}, \citenamefont {G{\"o}tz}, \citenamefont {Devoret},\
  and\ \citenamefont {Esteve}}]{Joyez99}%
  \BibitemOpen
  \bibfield  {author} {\bibinfo {author} {\bibfnamefont {P.}~\bibnamefont
  {Joyez}}, \bibinfo {author} {\bibfnamefont {D.}~\bibnamefont {Vion}},
  \bibinfo {author} {\bibfnamefont {M.}~\bibnamefont {G{\"o}tz}}, \bibinfo
  {author} {\bibfnamefont {M.}~\bibnamefont {Devoret}}, \ and\ \bibinfo
  {author} {\bibfnamefont {D.}~\bibnamefont {Esteve}},\ }\href {\doibase
  https://doi.org/10.1023/A:1007733009637} {\bibfield  {journal} {\bibinfo
  {journal} {Journal of superconductivity}\ }\textbf {\bibinfo {volume} {12}},\
  \bibinfo {pages} {757} (\bibinfo {year} {1999})}\BibitemShut {NoStop}%
\bibitem [{\citenamefont {Chauvin}(2005)}]{Chauvin05}%
  \BibitemOpen
  \bibfield  {author} {\bibinfo {author} {\bibfnamefont {M.}~\bibnamefont
  {Chauvin}},\ }\emph {\bibinfo {title} {The Josephson effect in atomic
  contacts}},\ \href@noop {} {Ph.D. thesis},\ \bibinfo  {school}
  {Universit{\'e} Pierre et Marie Curie-Paris VI} (\bibinfo {year}
  {2005})\BibitemShut {NoStop}%
\bibitem [{\citenamefont {Zellekens}\ \emph {et~al.}(2020)\citenamefont
  {Zellekens}, \citenamefont {Deacon}, \citenamefont {Perla}, \citenamefont
  {Fonseka}, \citenamefont {M\"orstedt}, \citenamefont {Hindmarsh},
  \citenamefont {Bennemann}, \citenamefont {Lentz}, \citenamefont {Lepsa},
  \citenamefont {Sanchez}, \citenamefont {Gr\"utzmacher}, \citenamefont
  {Ishibashi},\ and\ \citenamefont {Sch\"apers}}]{Zellekens20a}%
  \BibitemOpen
  \bibfield  {author} {\bibinfo {author} {\bibfnamefont {P.}~\bibnamefont
  {Zellekens}}, \bibinfo {author} {\bibfnamefont {R.}~\bibnamefont {Deacon}},
  \bibinfo {author} {\bibfnamefont {P.}~\bibnamefont {Perla}}, \bibinfo
  {author} {\bibfnamefont {H.~A.}\ \bibnamefont {Fonseka}}, \bibinfo {author}
  {\bibfnamefont {T.}~\bibnamefont {M\"orstedt}}, \bibinfo {author}
  {\bibfnamefont {S.~A.}\ \bibnamefont {Hindmarsh}}, \bibinfo {author}
  {\bibfnamefont {B.}~\bibnamefont {Bennemann}}, \bibinfo {author}
  {\bibfnamefont {F.}~\bibnamefont {Lentz}}, \bibinfo {author} {\bibfnamefont
  {M.~I.}\ \bibnamefont {Lepsa}}, \bibinfo {author} {\bibfnamefont {A.~M.}\
  \bibnamefont {Sanchez}}, \bibinfo {author} {\bibfnamefont {D.}~\bibnamefont
  {Gr\"utzmacher}}, \bibinfo {author} {\bibfnamefont {K.}~\bibnamefont
  {Ishibashi}}, \ and\ \bibinfo {author} {\bibfnamefont {T.}~\bibnamefont
  {Sch\"apers}},\ }\href {\doibase 10.1103/PhysRevApplied.14.054019} {\bibfield
   {journal} {\bibinfo  {journal} {Phys. Rev. Applied}\ }\textbf {\bibinfo
  {volume} {14}},\ \bibinfo {pages} {054019} (\bibinfo {year}
  {2020})}\BibitemShut {NoStop}%
\bibitem [{\citenamefont {{Forschungszentrum J\"ulich GmbH}}(2017)}]{HNF17}%
  \BibitemOpen
  \bibfield  {author} {\bibinfo {author} {\bibnamefont {{Forschungszentrum
  J\"ulich GmbH}}},\ }\href {http://dx.doi.org/10.17815/jlsrf-3-158} {\bibfield
   {journal} {\bibinfo  {journal} {Journal of large-scale facilities}\ }\textbf
  {\bibinfo {volume} {3}},\ \bibinfo {pages} {A112} (\bibinfo {year}
  {2017})}\BibitemShut {NoStop}%
\end{thebibliography}
\end{document}